%% file: Roth-Field-Circuit-QED-arXiv.tex
\documentclass[journal]{IEEEtran}

\usepackage{setspace}
\usepackage{epsfig}  % for figures
\usepackage{graphicx}  % another package that works for figures
\usepackage{amsmath}  % for math spacing
\usepackage{amssymb}  % for math spacing
\usepackage{lscape}  % Useful for wide tables or figures.
\usepackage[justification=raggedright]{caption}	% makes captions ragged right - thanks to Bryce Lobdell
\usepackage[labelformat=simple]{subcaption}

\usepackage{color}
\usepackage{cite}
\usepackage{mathrsfs}
\usepackage{mathtools}
\usepackage{amsthm}
\usepackage{array}
\usepackage{booktabs}
\usepackage{bm}

\ifCLASSINFOpdf
\else
\fi

\begin{document}

\title{Circuit Quantum Electrodynamics: A New Look Toward Developing Full-Wave Numerical Models}

\author{Thomas E. Roth~\IEEEmembership{Member,~IEEE},
        and Weng C. Chew~\IEEEmembership{Life Fellow,~IEEE}% <-this % stops a space
\thanks{This work was supported by NSF ECCS 169195, a startup fund at Purdue University, and the Distinguished Professorship Grant at Purdue University.
	
Thomas E. Roth is with the School of Electrical and Computer Engineering, Purdue University, West Lafayette, IN 47907 USA. Weng C. Chew is with the Department of Electrical and Computer Engineering, University of Illinois at Urbana-Champaign, Urbana, IL 61801 USA and the School of Electrical and Computer Engineering, Purdue University, West Lafayette, IN 47907 USA (contact e-mail: rothte@purdue.edu).

This work has been submitted to the IEEE for possible publication. Copyright may be transferred without notice, after which this version may no longer be accessible.}% <-this % stops a space
}

%\markboth{Journal of \LaTeX\ Class Files,~Vol.~13, No.~9, September~2014}%
%{Shell \MakeLowercase{\textit{et al.}}: Bare Demo of IEEEtran.cls for Journals}

\maketitle

\begin{abstract}
Devices built using circuit quantum electrodynamics architectures are one of the most popular approaches currently being pursued to develop quantum information processing hardware. Although significant progress has been made over the previous two decades, there remain many technical issues limiting the performance of fabricated systems. Addressing these issues is made difficult by the absence of rigorous numerical modeling approaches. This work begins to address this issue by providing a new mathematical description of one of the most commonly used circuit quantum electrodynamics systems, a transmon qubit coupled to microwave transmission lines. Expressed in terms of three-dimensional vector fields, our new model is better suited to developing numerical solvers than the circuit element descriptions commonly used in the literature. We present details on the quantization of our new model, and derive quantum equations of motion for the coupled field-transmon system. These results can be used in developing full-wave numerical solvers in the future. To make this work more accessible to the engineering community, we assume only a limited amount of training in quantum physics and provide many background details throughout derivations.
\end{abstract}


\IEEEpeerreviewmaketitle

\input{intro}

\input{transmon-background2}
\input{field-quantization2}

\input{field-tr-correspondence2}
\input{field-based-transmon2}
\input{eom}
\input{conclusion}
%\appendices
%\input{appendix}

%\section*{Acknowledgment}

%The authors would like to acknowledge useful discussions with Tian Xia, Shu Chen, Thomas Roth, Carlos Salazar-Lazaro and Professor %Scott P. Carney.
%We would like to acknowledge the following funding sources: UIUC CAS, AF Sub RRI PO0539, NSF ECCS 1609195, Ansys Inc PO37497 and the %George and Ann Fisher Professorship.

\ifCLASSOPTIONcaptionsoff
  \newpage
\fi

\bibliographystyle{IEEEtran}
% Put references in BibTeX format in thesisrefs.bib.
\bibliography{CEM_bib}


%\newpage



%\vfill


\end{document}

%% file: intro.tex
\section{Introduction}
\label{sec:intro}
\IEEEPARstart{C}{ircuit} quantum electrodynamics (QED) architectures are one of the leading candidates currently being pursued to develop quantum information processing devices \cite{blais2004cavity,blais2007quantum,gu2017microwave}, such as quantum simulators, gate-based quantum computers, and single photon sources \cite{ma2019dissipatively,arute2019quantum,zhou2020tunable,houck2007generating}. These circuit QED devices are most often formed by embedding superconducting Josephson junctions into planar microwave circuitry (also made from superconductors), such as coplanar waveguides \cite{blais2004cavity}. Doing this, planar ``on-chip'' realizations of many cavity QED and quantum optics concepts can be implemented at microwave frequencies. This provides a pathway to leveraging the fundamental light-matter interactions necessary to create and process quantum information in these new architectures \cite{gu2017microwave}. To achieve this, circuit QED systems implement ``artificial atoms'' (also frequently called qubits) through various configurations of Josephson junctions that are then coupled to coplanar waveguide resonators \cite{blais2004cavity,blais2007quantum,gu2017microwave}.

Circuit QED systems have garnered a high degree of interest in large part because of the achievable strength of light-matter coupling and the engineering control that is possible with these systems. The unique aspects of using artificial atoms formed by Josephson junctions allows for them to achieve substantially higher coupling strengths to electromagnetic fields than what is possible with natural atomic systems \cite{devoret2007circuit}. As a result, circuit QED systems have been able to achieve some of the highest levels of light-matter coupling strengths seen in any physical system to date, providing an avenue to explore and harness untapped areas of physics \cite{kockum2019ultrastrong}.

In addition to the strong coupling, circuit QED systems also provide a much higher degree of engineering control than is typically possible with natural atoms or ions. This is because many artificial atoms can be designed to have different desirable features by assembling Josephson junctions in various topologies \cite{gu2017microwave,vion2002manipulating,manucharyan2009fluxonium,koch2007charge}. Further, the operating characteristics of these artificial atoms can be tuned \textit{in situ} by applying various biases, such as voltages, currents, or magnetic fluxes \cite{gu2017microwave}. This allows dynamic reconfiguration of the artificial atom, opening possibilities for device designs that are not feasible using fixed systems like natural atoms or ions \cite{gu2017microwave}.  Finally, many aspects of the fabrication processes for these systems are mature due to their overlap with established semiconductor fabrication technologies \cite{gu2017microwave}.

Although there have been many successes with circuit QED systems to date (e.g., achieving ``quantum supremacy'' \cite{arute2019quantum}), a substantial amount of progress is still needed to truly unlock the potential of these systems. One area that could help accelerate the maturation of these technologies is the development of rigorous numerical modeling methods. Current models used in the physics community incorporate numerous approximations to simplify them to the point that they can be solved using semi-analytical approaches to build intuition about the physics \cite{blais2004cavity,blais2007quantum,koch2007charge,gu2017microwave,vool2017introduction,langford2013circuit,girvin2011circuit}. For instance, in almost all circuit QED studies, the electromagnetic aspects of the system are represented as simple combinations of lumped element LC circuits. Although this is appropriate and useful for building intuition, performing the engineering design and optimization of a practical device requires a level of precision that is not possible with this kind of \textit{circuit-based} description.

Instead, models that retain the full three-dimensional vector representation of the electromagnetic aspects of these circuit QED devices are needed. With these \textit{field-based} models, accurate full-wave numerical methods can begin to be formulated. These numerical methods can then be used to enable studies on the engineering optimization of circuit QED devices. Unfortunately, to the authors' best knowledge, this kind of detailed field-based description is not available in the literature. 

To address this issue, we present details in this work on the desired field-based framework for circuit QED systems and show how it can be used to derive the more commonly used circuit-based models found in the physics literature. To make the discussion more concrete, we focus on developing a field-based model for the transmon qubit \cite{koch2007charge}, which is one of the most widely used qubits in modern circuit QED systems \cite{ma2019dissipatively,arute2019quantum,zhou2020tunable,houck2007generating}. Similar procedures to those shown in this work can be applied to develop field-based models for other commonly used artificial atoms.

In an effort to make this work accessible to the engineering community, we provide many details in the derivations. We also assume only a limited amount of background knowledge in quantum physics at the level of \cite{chew2016quantum,chew2016quantum2,chew2021qme-made-simple}. Although circuit QED uses superconducting qubits, a minimal knowledge of superconductivity is needed to understand the general physics of these systems. Introductions to superconductivity in the context of circuit QED can be found in \cite{girvin2011circuit,langford2013circuit,vool2017introduction}. 

The remainder of this work is organized in the following way. In Section \ref{sec:transmon-background}, we review basic details of circuit QED systems using transmon qubits and introduce the field-based Hamiltonian developed in this work. Following this, we discuss quantization procedures in Section \ref{sec:field-quantization} for the electromagnetic fields that are devised specifically for developing numerical methods. Next, Section \ref{sec:field-to-circuit} presents details on how field-based descriptions can be converted into a transmission line formalism. Using these details, Section \ref{sec:field-transmon-hamiltonian} shows how the field-based Hamiltonian for circuit QED systems is consistent with the circuit-based descriptions found in the literature. With an appropriate Hamiltonian developed, Section \ref{sec:eom} derives the field-based quantum equations of motion that can be used for formulating new numerical modeling strategies. Finally, we present conclusions on this work in Section \ref{sec:conclusion}.

%% file: transmon-background2.tex
\section{Circuit QED Preliminaries}
\label{sec:transmon-background}
To support the development of the field-based description of circuit QED systems, it will first be necessary to review a few properties of these systems. We begin this by discussing the basic physical properties of the transmon qubit in Section \ref{subsec:transmon-basic-physics}. Following this, we discuss in Section \ref{subsec:transmon-coupling} how the coupling of a transmon qubit to a transmission line is typically handled using a circuit theory description. Finally, in Section \ref{subsec:field-based} we briefly introduce the field-based description of the transmon qubit coupled to a transmission line structure. We will demonstrate the consistency of the field- and circuit-based descriptions of this system in Section \ref{sec:field-transmon-hamiltonian} after developing the necessary tools in the intervening sections. 

For readers interested in a more complete description of the transmon qubit, we refer them to the seminal work of \cite{koch2007charge} that presents an in-depth theoretical analysis. More details on the derivation of the typically used Hamiltonians introduced in this section can be found in \cite{vool2017introduction,girvin2011circuit,langford2013circuit,kockum2019quantum}. We focus our discussions on the basic physics of the Hamiltonians to provide an intuitive understanding only. 

\subsection{Basic Physical Properties of a Transmon}
\label{subsec:transmon-basic-physics}
Typically, quantum effects are only observable at microscopic levels due to the fragility of individual quantum states. To observe quantum behavior at a macroscopic level (e.g., on the size of circuit components), a strong degree of coherence between the individual microscopic quantum systems must be achieved \cite{girvin2011circuit}. One avenue for this to occur is in superconductors cooled to extremely low temperatures (on the order of 10 mK) \cite{kockum2019quantum}. At these temperatures, electrons in the superconductor tend to become bound to each other as Cooper pairs \cite{vool2017introduction,girvin2011circuit,langford2013circuit}. These Cooper pairs exhibit bosonic properties and become the charge carriers of the superconducting system. Importantly, they have the necessary degree of coherence over large length scales to make observing macroscopic quantum behavior possible. Circuit QED systems interact with these macroscopic quantum states using microwave photons and other circuitry \cite{gu2017microwave}.

One of the earliest qubits used to observe macroscopic quantum behavior in circuit QED systems was the Cooper pair box (CPB) \cite{nakamura1999coherent}, which can be viewed as a predecessor to the transmon. The traditional CPB is formed by a thin insulative gap (on the order of a nm thick) that connects a superconducting ``island'' and a superconducting ``reservoir'' \cite{kockum2019quantum}. The superconductor-insulator-superconductor ``sandwich'' formed between the island and reservoir is known as a Jospehson junction, and has the property that Cooper pairs may tunnel through the junction without requiring an applied voltage \cite{tafuri2019introductory}. For the basic CPB, the island is not directly connected to other circuitry, while the reservoir can be in contact with external circuit components (if desired). Since the superconducting island is isolated from other circuitry, the CPB is very sensitive to the number of Cooper pairs that have tunneled through the Josephson junction. Due to this sensitivity, the CPB is also often referred to as a charge qubit \cite{kockum2019quantum}.

As is common in quantum physics, it is desirable to consider a Hamiltonian mechanics description of the CPB \cite{chew2016quantum,chew2016quantum2,chew2021qme-made-simple}. For an isolated system, this amounts to expressing the total energy of the Josephson junction in terms of \textit{canonical conjugate variables}. These conjugate variables vary with respect to each other in a manner to ensure that the total energy of the system is conserved \cite{chew2021qme-made-simple}. For the CPB system, the canonical conjugate variables are the Cooper pair density difference $n$ and the Josephson phase $\varphi$. Initially considering the classical case, these variables are real-valued deterministic numbers.

For this case, $n$ is the net density of Cooper pairs that have tunneled through the Josephson junction relative to an equilibrium level \cite{vool2017introduction,girvin2011circuit,langford2013circuit}. Due to its relationship to a microscopic theory of superconductivity, $\varphi$ is more challenging to interpret \cite{langford2013circuit}. Briefly, there exists a long-range phase coherence for a collective description of all the Cooper pairs in a superconductor. As a result, the phase of the collective description becomes a meaningful variable to characterize the momentum of all Cooper pairs. This phase can then be related to a current flowing in the superconductor \cite{langford2013circuit}. For a Josephson junction, the phase difference of the two superconductors is important, and is denoted as $\varphi$ \cite{vool2017introduction,girvin2011circuit,langford2013circuit}.

The Hamiltonian of the CPB can be found by considering the total energy of the junction in terms of an effective capacitance and inductance expressed with $n$ and $\varphi$. The capacitance is due to the ``parallel plate'' configuration of the junction. The energy is found by first noting that $2en = Q$, where $Q$ is the total charge ``stored'' in the junction capacitance and $2e$ is the charge of a Cooper pair. Now, considering that the single electron charging energy of a capacitor is $E_C = e^2/2C$, the total capacitive energy of the junction capacitance, $Q^2/2C$, can be written as $4 E_C n^2$ \cite{girvin2011circuit}.

The inductance of the Josephson junction is more complicated to understand because it involves the tunneling physics. However, for the discussion here it only needs to be known that the supercurrent flowing through the junction is $I=I_c \sin{\varphi}$, where $I_c$ is the critical current of the junction. Since $\partial_t \varphi$ can be related to the voltage drop over the junction, this current-phase relationship can be used to derive an effective Josephson inductance that is proportional to $1/\cos{\varphi}$ \cite{girvin2011circuit}. Overall, the energy associated with the inductance is $-E_J\cos{\varphi}$, where $E_J$ is the Josephson energy.

Combining the results for the effective capacitive and inductive energy, the Hamiltonian for the CPB is
\begin{align}
H_T = 4E_C n^2 - E_J \cos\varphi.
\label{eq:classical-JJ-Hamiltonian}
\end{align}
This Hamiltonian can be viewed as being equivalent to a linear capacitor in parallel with a nonlinear inductor. It is this nonlinear inductance that allows Josephson junctions to form qubits. Without the nonlinearity, the energy levels of a quantized form of (\ref{eq:classical-JJ-Hamiltonian}) would be evenly spaced, making it impossible to selectively target a single pair of energy levels to perform qubit operations. If needed, the equations of motion for the CPB can be derived from (\ref{eq:classical-JJ-Hamiltonian}) using Hamilton's equations \cite{chew2016quantum,chew2016quantum2,chew2021qme-made-simple}.

\begin{figure}[t]
	\centering
	\begin{subfigure}[t]{0.35\linewidth}
		\includegraphics[width=\textwidth]{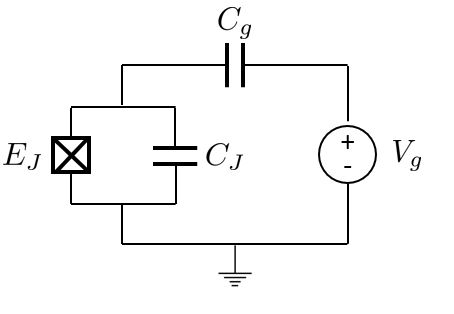}
		\caption{}
		\label{subfig:basic_cpb}
	\end{subfigure}
	\begin{subfigure}[t]{0.435\linewidth}
		\includegraphics[width=\textwidth]{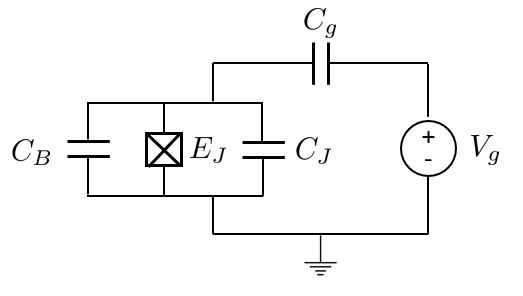}
		\caption{}
		\label{subfig:transmon}
	\end{subfigure}
	\caption{Circuit schematics for (a) a traditional CPB and (b) a transmon. A Josephson junction consists of a pure tunneling element (symbolized as a box with an ``X'' through it) in parallel with a small junction capacitance $C_J$.}
	\label{fig:cpb_schematics}
\end{figure}

Typically, it is desirable to be able to control the operating point of the CPB system. This can be done using a voltage source capacitively coupled to the superconducting island. This induces a ``background'' Cooper pair density that has tunneled onto the island, denoted by $n_g$ (this is also often referred to as the offset charge) \cite{kockum2019quantum}. The basic circuit diagram of this qubit is shown in Fig. \ref{subfig:basic_cpb}, where $V_g$ is an applied voltage bias capacitively coupled to the CPB through $C_g$. For this system, the Hamiltonian is modified to be  
\begin{align}
H_T = 4E_C (n-n_g)^2 - E_J \cos\varphi.
\label{eq:classical-JJ-Hamiltonian2}
\end{align}

The system can now be quantized by elevating the canonical conjugate variables to be non-commuting quantum operators. In particular, we now characterize the CPB with the quantum operators $\hat{n}$ and $\hat{\varphi}$ that have commutation relation \cite{vool2017introduction}
\begin{align}
[\hat{\varphi},\hat{n}] = i.
\end{align}
Combined with a complex-valued quantum state function, these quantum operators take on a statistical interpretation and share an uncertainty principle relationship \cite{chew2021qme-made-simple}. As a result, measurements of observables associated with these quantum operators (e.g., the number of Cooper pairs that have tunneled through the junction) become random variables with means and variances dictated by the laws of quantum mechanics \cite{chew2021qme-made-simple}. The combination of these properties allows for the non-classical interference between possible states of a system, which is key to quantum information processing.

Now, one advantage of determining the classical Hamiltonian of the CPB system is that the quantum Hamiltonian follows easily from it \cite{chew2021qme-made-simple}. In particular, the quantum Hamiltonian is \cite{kockum2019quantum,koch2007charge}
\begin{align}
\hat{H}_T = 4 E_C (\hat{n}-n_g)^2 - E_J \cos \hat{\varphi}.
\label{eq:isolated_cpb_hamiltonian}
\end{align}
Note that $n_g$ remains a classical variable that describes the offset charge induced by the applied DC voltage.

The different terms in (\ref{eq:isolated_cpb_hamiltonian}) take on similar physical meaning to the classical case of (\ref{eq:classical-JJ-Hamiltonian2}). However, the second term can be expressed in the charge basis as
\begin{align}
-E_J\cos\hat{\varphi} = \frac{E_J}{2} \sum_N \big[ |N\rangle\langle N+1 | + | N+1\rangle\langle N |   \big],
\label{eq:tunneling-hamiltonian}
\end{align}
where $|N \rangle$ is an eigenstate of $\hat{n}$ with eigenvalue $N$ \cite{vool2017introduction}. This eigenvalue is a discrete number that counts how many Cooper pairs have tunneled through the junction. Considering this, we see that the form of the effective inductive energy given in (\ref{eq:tunneling-hamiltonian}) clearly shows the tunneling physics. 

Unfortunately, the CPB is very sensitive to charge fluctuations (i.e., noise) that appear from a variety of sources in $n_g$ \cite{koch2007charge}. This sensitivity prevents the CPB from being applicable to scalable quantum information processing systems. To address this issue, the transmon qubit was introduced as an ``optimized CPB''. The differences between these qubits are most easily understood in terms of the ratio of $E_J$ to $E_C$ where the devices are designed to operate. In traditional CPBs, $E_J/E_C \ll 1$, due to the small capacitance $C_J$ naturally provided by the Josephson junction. The transmon qubit is designed to operate like a CPB, but with $E_J/E_C \gg 1$. 

\begin{figure}[t]
	\centering
	\includegraphics[width=1.0\linewidth]{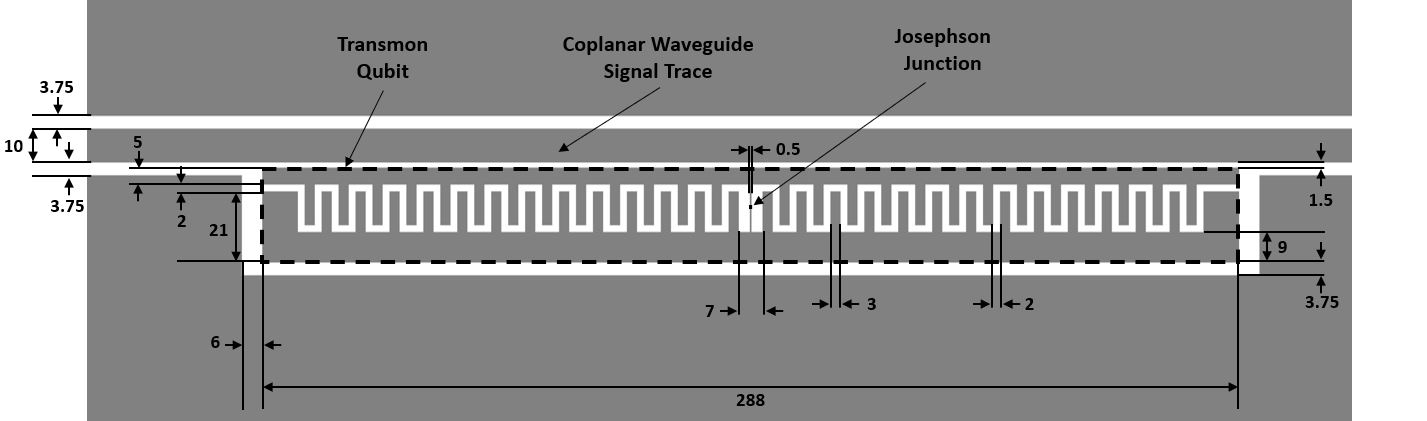}
	\caption{Schematic of a transmon qubit coupled to a coplanar waveguide transmission line. Dimensions (in $\mu$m) are based on the transmon qubit used in \cite{houck2007generating}.}
	\label{fig:physical_transmon}
\end{figure}

This is achieved by dramatically decreasing $E_C$ by adding a large shunting capacitance $C_B$ around the CPB, as shown in Fig. \ref{subfig:transmon}. This capacitance is often implemented by forming interdigital capacitors between two superconducting islands that are connected by a Josephson junction, as shown in Fig. \ref{fig:physical_transmon}. Although the physical implementation of the transmon is different from the CPB, the same Hamiltonian of (\ref{eq:isolated_cpb_hamiltonian}) describes its behavior \cite{koch2007charge}. However, because $E_C \ll E_J$ the transmon becomes insensitive to fluctuations in $n_g$, making it useful for scalable quantum information processing systems \cite{koch2007charge}. 

\subsection{Coupling the Transmon to a Transmission Line Resonator}
\label{subsec:transmon-coupling}
A completely isolated qubit cannot be controlled, and so is of little use. Circuit QED systems address this by coupling a qubit to transmission line structures that can be used to apply microwave drive pulses to control and read out the qubit's state, as well as to interface separated qubits to implement qubit-qubit interactions \cite{blais2004cavity,blais2007quantum,gu2017microwave}. For transmons, capacitive coupling to transmission lines is typical \cite{koch2007charge}. One early strategy using a coplanar waveguide is shown in Fig. \ref{fig:physical_transmon}. More recently, other strategies have emerged to improve the interconnectivity to the transmon \cite{barends2014superconducting}. However, analyzing these newer implementations follows the analysis of the coupling shown in Fig. \ref{fig:physical_transmon}, so only the simpler case will be considered here. 

The interaction between the transmon and transmission line can be described quantum mechanically in a number of ways. However, it is usually convenient to express the interaction in terms of $\hat{n}$ and a transmission line voltage operator  \cite{koch2007charge}. Often, the transmission line the transmon is coupled to is a resonator. As a result, the voltage operator can be conveniently written in terms of the modes of the resonator \cite{blais2004cavity}. These modes are the one-dimensional sinusoidal functions used to describe the spatial dependence of a resonator's voltage and current in microwave engineering \cite{pozar2009microwave}. This spatial dependence is integrated out of the Hamiltonian to arrive at a circuit-based (i.e., lumped element) description of the interaction.

Following this process, the resulting Hamiltonian that describes a single transmon qubit coupled to a single mode of a transmission line resonator is
\begin{align}
\hat{H} = \hat{H}_T + \hat{H}_R + 2e\beta V_g \hat{V}_r \hat{n},
\label{eq:coupled-transmon}
\end{align}
\begin{align}
\hat{H}_R = \frac{1}{2}\big[ L_r \hat{I}_r^2 + C_r \hat{V}_r^2   \big],
\label{eq:free-resonator}
\end{align}
where $\hat{I}_r$ and $\hat{V}_r$ are the \textit{integrated} resonator voltage and current (i.e., the spatial variation of the mode has been integrated out) \cite{blais2004cavity,koch2007charge}. The Hamiltonian in (\ref{eq:free-resonator}) is the \textit{free resonator Hamiltonian}, i.e., it describes the total energy of the uncoupled resonator mode (and can be viewed as being equivalent to an LC tank circuit). The final term in (\ref{eq:coupled-transmon}) represents the interaction between the transmon and resonator, in terms of the resonator voltage and transmon charge operators.  

Other terms in (\ref{eq:coupled-transmon}) and (\ref{eq:free-resonator}) include the magnitude of the resonator voltage at the location of the transmon $V_g$ and $\beta = C_g/(C_g+C_B)$. The latter term represents a voltage divider to capture the portion of the resonator voltage applied to the transmon within a lumped element approximation (note $C_J$ has been absorbed into the much larger capacitance $C_B$ in $\beta$). Further, $L_r = \ell L$ and $C_r = \ell C$, where $\ell$ is the length of the resonator and $L$ and $C$ are the per-unit-length inductance and capacitance of the transmission line. These terms arise from the normalization and subsequent spatial integration of the resonator voltage and current modes. Extending this model to contain multiple resonator modes is in principle quite simple, and will be considered later in this work.

Depending on the numerical model being developed, it may or may not be desirable to keep the interdigital capacitor in the geometric description of the system. When it is desirable to explicitly model the interdigital capacitor, $\beta$ should be omitted from equations since the ``voltage divider'' it represents would already be accounted for. To allow the equations in this work to be used in either situation, we keep $\beta$ in all equations.

Most circuit QED studies work with the transmission line operators expressed in terms of bosonic ladder operators. In particular, each mode of the transmission line is viewed as an independent quantum harmonic oscillator characterized by bosonic ladder operators \cite{chew2016quantum2}. The integrated resonator voltage and current operators in terms of the ladder operators are
\begin{align}
\hat{V}_r = \sqrt{\frac{\hbar \omega_r}{2C_r}} (\hat{a}+\hat{a}^\dagger),
\label{eq:total-v}
\end{align}
\begin{align}
\hat{I}_r = -i\sqrt{\frac{\hbar \omega_r}{2L_r}} (\hat{a}-\hat{a}^\dagger),
\label{eq:total-i}
\end{align}
where $\omega_r$ is the resonant frequency of the mode \cite{blais2004cavity,koch2007charge}. The ladder operators satisfy the bosonic commutation relation,
\begin{align}
[\hat{a},\hat{a}^\dagger] = 1.
\end{align}
Using properties of the ladder operators, the Hamiltonian in (\ref{eq:coupled-transmon}) can be expressed as
\begin{align}
\hat{H} = \hat{H}_T + \hbar\omega_r \hat{a}^\dagger\hat{a}  + 2e\beta V^\mathrm{rms}_g \hat{n} (\hat{a}+\hat{a}^\dagger),
\label{eq:coupled-transmon2}
\end{align}
where $V^\mathrm{rms}_g = V_g \sqrt{\hbar\omega_r/2C_r}$ and the zero point energy of the transmission line resonator has been adjusted to remove constant terms. 

The isolated transmon Hamiltonian of (\ref{eq:isolated_cpb_hamiltonian}) can be diagonalized exactly \cite{koch2007charge}. Using these eigenstates (denoted as $|j\rangle$), the complete system Hamiltonian of (\ref{eq:coupled-transmon2}) can be rewritten as
\begin{multline}
\hat{H} = \sum_j \hbar\omega_j |j\rangle\langle j | + \hbar\omega_r \hat{a}^\dagger\hat{a} \\ + 2e\beta V_g^\mathrm{rms} \sum_{i,j} \langle i | \hat{n} | j \rangle \, |i\rangle \langle j | (\hat{a}+\hat{a}^\dagger) ,
\label{eq:coupled-transmon3}
\end{multline}
where $\omega_j$ is the eigenvalue associated with the $j$th transmon eigenstate. A more complete description of these eigenstates can be found in \cite{koch2007charge}. For this work, the essential property is that transitions between two transmon eigenstates corresponds to a high probability event of some number of Cooper pairs tunneling through the Josephson junction.

In the transmon operating regime, (\ref{eq:coupled-transmon3}) can be further simplified because it is safe to assume that the $\hat{n}$ operator only couples nearest neighbor transmon eigenstates. Hence, (\ref{eq:coupled-transmon3}) can be rewritten as
\begin{multline}
\hat{H} = \sum_j \hbar\omega_j |j\rangle\langle j | + \hbar\omega_r \hat{a}^\dagger\hat{a}  + 2e\beta V_g^\mathrm{rms} \times \\ \sum_{i} \langle i | \hat{n} | i+1 \rangle  \, \big(|i\rangle \langle i+1 | + |i+1\rangle \langle i | \big) \big(\hat{a}+\hat{a}^\dagger \big) ,
\label{eq:coupled-transmon4}
\end{multline}
where the fact that $\langle i | \hat{n} | i+1 \rangle = \langle i+1 | \hat{n} | i \rangle $ has also been used \cite{koch2007charge}. This is the circuit-based Hamiltonian that is typically used as a starting point for many circuit QED studies. 

\subsection{Field-Based Description of the Transmon System}
\label{subsec:field-based}
With the physics of the transmon understood, we can now briefly introduce the field-based description of a circuit QED system using a transmon. In particular, the postulated field-transmon system Hamiltonian is
\begin{align}
\hat{H} =  \hat{H}_T + \hat{H}_F - \iiint  \hat{\mathbf{E}} \cdot \partial_t^{-1} \hat{\mathbf{J}}_t  d\mathbf{r}.
\label{eq:field-transmon-hamiltonian1}
\end{align}
In (\ref{eq:field-transmon-hamiltonian1}), $\hat{H}_T$ is the transmon Hamiltonian given in (\ref{eq:isolated_cpb_hamiltonian}),
\begin{align}
\hat{H}_F = \frac{1}{2}  \iiint \big( \epsilon \hat{\mathbf{E}}^2 + \mu \hat{\mathbf{H}}^2 \big) d\mathbf{r}
\end{align}
is the free field Hamiltonian consisting of the electric and magnetic field operators $\hat{\mathbf{E}}$ and $\hat{\mathbf{H}}$, and the final term is the interaction between $\hat{\mathbf{E} }$ and a transmon current density operator $\hat{\mathbf{J}}_t$. The transmon current density operator is
\begin{align}
\hat{\mathbf{J}}_t = -2e \beta \mathbf{d} \delta(z-z_0) \partial_t\hat{n} ,
\label{eq:transmon-current-operator1}
\end{align}
where $\hat{n}$ is the standard Josephson junction charge operator. We use the awkward notation of $\partial_t^{-1}\hat{\mathbf{J}}_t$ in (\ref{eq:field-transmon-hamiltonian1}) since it will be convenient to use a current density operator when deriving equations of motion in Section \ref{sec:eom}. In (\ref{eq:transmon-current-operator1}), $\mathbf{d}$ is a vector parameterizing the integration path taken to define the voltage of the transmission line at the location of the transmon. Further, the $z$-axis has been identified as the longitudinal direction of the transmission line in the region local to the transmon for notational simplicity.

Before continuing, it is worth commenting on the interpretation of $\hat{\mathbf{J}}_t$. Within the transmon basis, (\ref{eq:transmon-current-operator1}) becomes
\begin{multline}
\hat{\mathbf{J}}_t = -2e\beta\mathbf{d} \delta(z-z_0) \\ \times\sum_j  \langle j |  \hat{n} | j + 1\rangle \partial_t\big[  |j\rangle\langle j+1| +  |j+1\rangle\langle j| \big],
\label{eq:transmon-current-operator3}
\end{multline}
after applying the result that only nearest neighbor states couple for a transmon \cite{koch2007charge}. From this, we see that $\hat{\mathbf{J}}_t$ involves transitions between different transmon eigenstates. These transitions correspond to a high probability event of Cooper pairs tunneling through the Josephson junction. This tunneling produces a current, making the designation of (\ref{eq:transmon-current-operator1}) as a current density reasonable.

Although the physics of (\ref{eq:field-transmon-hamiltonian1}) is fairly intuitive, we will need to develop a number of tools in the following sections to demonstrate its consistency with the circuit-based description of (\ref{eq:coupled-transmon}). This will first require a careful look at the quantization of electromagnetic fields for circuit QED systems in Section \ref{sec:field-quantization}, followed by establishing a correspondence between field and transmission line representations of quantum operators in Section \ref{sec:field-to-circuit}. We will then demonstrate the consistency of (\ref{eq:field-transmon-hamiltonian1}) and (\ref{eq:coupled-transmon}) in Section \ref{sec:field-transmon-hamiltonian}.

%% file: field-quantization2.tex
\section{Field Quantization for Circuit QED Systems}
\label{sec:field-quantization}
\begin{figure}[t]
	\centering
	\includegraphics[width=0.9\linewidth]{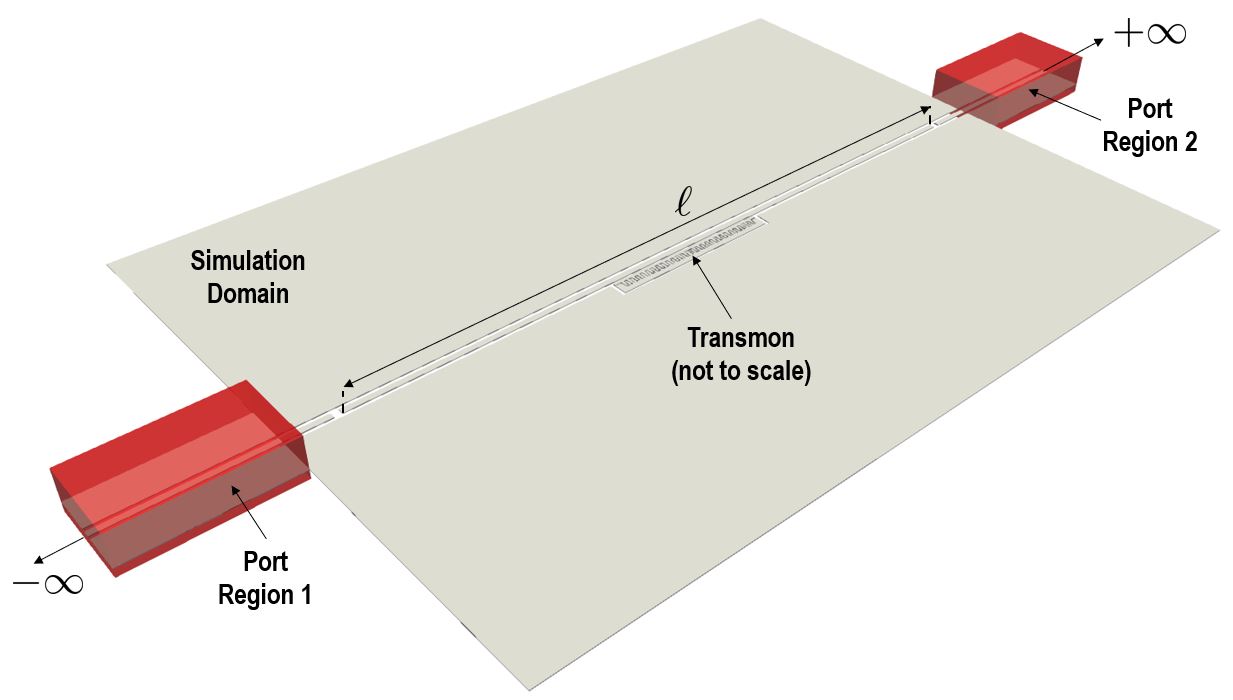}
	\caption{Example of region definitions for a simple circuit QED system composed of a transmon qubit (not to scale) coupled to a coplanar waveguide resonator of length $\ell$. The red regions are fictitious boundaries for the port regions. The dielectric substrate of the system is not shown.}
	\label{fig:region-illustration}
\end{figure}

We present two approaches for quantizing the electromagnetic field in systems containing inhomogeneous, lossless, and non-dispersive dielectric and perfectly conducting regions. To simplify the notation, we assume there are no magnetic materials present. The two quantization approaches are relevant for developing different numerical methods. The first approach, discussed in Section \ref{subsec:modes-of-the-universe}, follows a standard mode decomposition quantization process \cite{chew2016quantum2,walls2007quantum,gerry2005introductory,viviescas2003field}. We will refer to this as the \textit{modes-of-the-universe} quantization approach, since the spatial modes used extend across all space \cite{viviescas2003field}. For many numerical modeling approaches, it is convenient to consider a finite-sized \textit{simulation domain} with a number of semi-infinite \textit{port regions} attached to it, as illustrated in Fig. \ref{fig:region-illustration}. This is not compatible with the modes-of-the-universe approach, and so, a different quantization approach will be discussed in Section \ref{subsec:projector-quantization} for this case. 

Both quantization methods discussed are performed within the framework of macroscopic QED \cite{scheel2008macroscopic}. The key aspect of this is that a microscopic description of a lossless, non-dispersive dielectric medium is unnecessary. As a result, macroscopic permittivities and permeabilities can be directly used in a quantum description of the electromagnetic fields. Accounting for dispersive or lossy media is more complicated, and is outside the scope of this work \cite{wei2018dissipative,milonni1995field,gruner1996green}.

In some situations, requiring a mode decomposition description can be inconvenient. For instance, this can occur when dealing with certain kinds of nonlinearities. If this is the case, an energy conservation argument can be applied to quantize the electromagnetic field directly in coordinate space \cite{chew2021qme-made-simple}.

\subsection{Modes-of-the-Universe Quantization}
\label{subsec:modes-of-the-universe}
Quantizing the electromagnetic field using a modes-of-the-universe framework is one of the most common quantization approaches. Introductory reviews can be found in \cite{chew2016quantum2,gerry2005introductory,walls2007quantum}. This technique is often simply referred to as a mode decomposition approach. The longer terminology will be used in this work to differentiate it from the quantization approach discussed in Section \ref{subsec:projector-quantization}.

The first step of the modes-of-the-universe approach is to use separation of variables to write the electric field as
\begin{align}
\mathbf{E}(\mathbf{r},t) = \sum_k \sqrt{\frac{\omega_k}{2\epsilon_0}} \big( q_k(t) \mathbf{E}_k(\mathbf{r}) + q_k^*(t) \mathbf{E}^*_k(\mathbf{r}) \big).
\end{align}
For initial simplicity, a discrete summation of modes is assumed. A continuum of modes will be considered as part of the quantization procedure discussed in Section \ref{subsec:projector-quantization}. Inserting this representation into
\begin{align}
\nabla\times\nabla\times\mathbf{E} + \mu\epsilon \partial_t^2 \mathbf{E} = 0
\end{align}
yields two separated equations for each mode, given by
\begin{align}
\partial_t^2 q_k(t) = -\omega_k^2 q_k(t),
\label{eq:mode_time}
\end{align}
\begin{align}
\nabla\times\nabla\times\mathbf{E}_k(\mathbf{r}) - \mu\epsilon \omega_k^2 \mathbf{E}_k(\mathbf{r}) = 0.
\label{eq:field-eig-def}
\end{align}
The complex conjugates of $q_k$ and $\mathbf{E}_k$ also obey (\ref{eq:mode_time}) and (\ref{eq:field-eig-def}), respectively, since $\omega_k$, $\epsilon$, and $\mu$ are all real. Considering (\ref{eq:mode_time}), it is easily seen that the time dependence of these modes will be $\exp(\pm i \omega_k t)$.  

To simplify the analysis, it is further required that the field modes are orthonormal such that
\begin{align}
\iiint \epsilon_r(\mathbf{r}) \mathbf{E}^*_{k_1}(\mathbf{r}) \cdot \mathbf{E}_{k_2}(\mathbf{r}) d\mathbf{r} = \delta_{k_1,k_2},
\label{eq:orthonormal-def}
\end{align}
where $\delta_{k_1,k_2}$ is the Kronecker delta function. Later, it will be useful to explicitly consider the normalization of the field modes. Hence, we will often write the modes as
\begin{align}
\mathbf{E}_k = \frac{1}{\sqrt{N_{u,k}}} \mathbf{u}_k(\mathbf{r}),
\end{align}
where
\begin{align}
N_{u,k} = \iiint \epsilon_r(\mathbf{r}) \mathbf{u}_k^*(\mathbf{r}) \cdot \mathbf{u}_k(\mathbf{r}) dV.
\label{eq:original-enorm}
\end{align}

Similarly, the magnetic field can be written as
\begin{align}
\mathbf{H}(\mathbf{r},t) = \sum_k \sqrt{\frac{\omega_k}{2\mu_0}} \big( p_k(t) \mathbf{H}_k(\mathbf{r}) +  p_k^*(t) \mathbf{H}^*_k(\mathbf{r}) \big),
\end{align}
where
\begin{align}
\mathbf{H}_k = \frac{1}{\sqrt{N_{v,k}} } \mathbf{v}_k(\mathbf{r})
\end{align}
and the normalization for the magnetic field modes take on a similar form to (\ref{eq:original-enorm}). Although the magnetic field modes have been denoted by seemingly independent variables, i.e., $\mathbf{v}_k$ and $p_k$, they are related to the electric field variables $\mathbf{u}_k$ and $q_k$ according to Maxwell's equations. 

It should be noted that these expansions are valid for complex-valued spatial modes. In a closed region (e.g., a cavity), it is often advantageous to use real-valued spatial modes. For this situation, the field expansions become
\begin{align}
\mathbf{E}(\mathbf{r},t) = \sum_k \sqrt{\frac{\omega_k}{2\epsilon_0}} \big( q_k(t)  + q_k^*(t)  \big) \mathbf{E}_k(\mathbf{r}),
\end{align}
\begin{align}
\mathbf{H}(\mathbf{r},t) = -i\sum_k \sqrt{\frac{\omega_k}{2\mu_0}} \big( p_k(t)  - p_k^*(t)  \big) \mathbf{H}_k(\mathbf{r}).
\end{align}
It will be useful to use both real- and complex-valued spatial mode functions in this work. 

With the mode expansion defined, the Hamiltonian for the electromagnetic field system can be expanded in terms of these modes. The electromagnetic field Hamiltonian is equivalent to the total electromagnetic energy in a system, which is
\begin{align}
H_F = \iiint \frac{1}{2}\big( \epsilon |\mathbf{E}(\mathbf{r},t)|^2 + \mu |\mathbf{H}(\mathbf{r},t)|^2     \big) d\mathbf{r}.
\label{eq:free-field-cH}
\end{align}
Substituting in either the real- or complex-valued mode expansions and performing the spatial integrations, the Hamiltonian simplifies to
\begin{align}
H_F = \sum_k \frac{\omega_k}{2} \big( |q_k(t)|^2 + |p_k(t)|^2  \big).
\end{align}
This can be readily identified as a summation of Hamiltonians for uncoupled harmonic oscillators, or equivalently uncoupled LC resonant circuits \cite{chew2016quantum2}.

Hence, a canonical quantization process can now be performed by elevating the conjugate variables of each harmonic oscillator (i.e., the $q_k$ and $p_k$) to be quantum operators \cite{chew2016quantum2}. These operators obey the canonical commutation relation
\begin{align}
[\hat{q}_{k_1},\hat{p}_{k_2}] = i\hbar \delta_{k_1,k_2} .
\end{align}
These operators may be combined to form bosonic ladder operators for each quantum harmonic oscillator. This gives the annihilation operator as
\begin{align}
\hat{a}_k = \frac{1}{\sqrt{2\hbar}}(\hat{q}_k + i \hat{p}_k)
\label{eq:annihilation}
\end{align}
and the creation operator as
\begin{align}
\hat{a}^\dagger_k = \frac{1}{\sqrt{2\hbar}}(\hat{q}_k - i \hat{p}_k).
\end{align}
These operators satisfy the bosonic commutation relation
\begin{align}
[\hat{a}_{k_1},\hat{a}^\dagger_{k_2}] = \delta_{k_1,k_2} .
\label{eq:boson-commutation}
\end{align}

In terms of ladder operators, the field operators become
\begin{align}
\hat{\mathbf{E}}(\mathbf{r},t) = \sum_k N_{E,k} \big( \hat{a}_k(t)\mathbf{u}_k(\mathbf{r}) + \hat{a}_k^\dagger(t)\mathbf{u}^*_k(\mathbf{r}) \big) 
\label{eq:c-q-efield}
\end{align}
\begin{align}
\hat{\mathbf{H}}(\mathbf{r},t) = \sum_k N_{H,k} \big( \hat{a}_k(t)\mathbf{v}_k(\mathbf{r}) + \hat{a}_k^\dagger(t)\mathbf{v}^*_k(\mathbf{r}) \big)
\label{eq:c-q-hfield}
\end{align}
for complex-valued spatial modes. In (\ref{eq:c-q-efield}) and (\ref{eq:c-q-hfield}),
\begin{align}
N_{E,k} = \sqrt{\frac{\hbar \omega_k}{2\epsilon_0 N_{u,k} }}, \,\, N_{H,k} = \sqrt{\frac{\hbar \omega_k}{2\mu_0 N_{v,k} }}.
\label{eq:e-norm}
\end{align} 
Similarly, the real-valued spatial mode expansions give
\begin{align}
\hat{\mathbf{E}}(\mathbf{r},t) = \sum_k N_{E,k} \big( \hat{a}_k(t) + \hat{a}_k^\dagger(t) \big) \mathbf{u}_k(\mathbf{r}),
\label{eq:r-q-efield}
\end{align}
\begin{align}
\hat{\mathbf{H}}(\mathbf{r},t) = -i\sum_k N_{H,k} \big( \hat{a}_k(t) - \hat{a}_k^\dagger(t) \big) \mathbf{v}_k(\mathbf{r}).
\label{eq:r-q-hfield}
\end{align}
The quantum field Hamiltonian becomes
\begin{align}
\hat{H}_{F} = \iiint \frac{1}{2} \big( \epsilon \hat{\mathbf{E}}^2 + \mu \hat{\mathbf{H}}^2 \big) d\mathbf{r},
\label{eq:free-field-qH}
\end{align}
which after spatial integration and adjusting the zero point energy can be written in a ``diagonalized'' form in terms of the ladder operators as
\begin{align}
\hat{H}_{F} = \sum_k \hbar \omega_k \hat{a}_k^\dagger \hat{a}_k.
\label{eq:universe-hamiltonian}
\end{align}
As expected, the final Hamiltonian closely matches the isolated circuit portion of the Hamiltonian given in (\ref{eq:coupled-transmon4}). 

\subsection{Projector-Based Quantization}
\label{subsec:projector-quantization}
The quantization process in Section \ref{subsec:modes-of-the-universe} is not desirable when port regions like those shown in Fig. \ref{fig:region-illustration} are needed to model a circuit QED system. The issue is that the modes-of-the-universe approach makes no distinction between modes that are associated with ``internal'' dynamics (e.g., that of a transmon coupled to a resonator) and ``external'' modes leaving the device (e.g., modes entering or exiting a device via a port). As a result, a different quantization procedure that allows for the modes in the various regions to be independently worked with is of more interest. 

One approach to do this is the Feshbach projector technique as applied to quantum optics \cite{viviescas2003field,viviescas2004quantum}. This approach defines a set of projection operators to isolate the behavior in the various regions of the problem. The eigenvalue problem from the modes-of-the-universe approach is then projected into a set of eigenvalue problems for the various regions being considered. The hermiticity of the projected eigenvalue problems can be maintained by selecting complementary boundary conditions to apply at the interfaces between regions \cite{viviescas2003field}. As a result, a complete set of orthogonal eigenmodes can be found for each region. These various modes can then be quantized and coupled to each other.

For modeling a circuit QED system, a natural decomposition would have one projector cover the simulation domain and another set of complementary projectors cover the infinitely long transmission lines that model ports, as illustrated in Fig. \ref{fig:region-illustration}. For clarity, the region of the simulation domain will be denoted by $\mathcal{Q}$ and the various port regions by $\mathcal{P}_p$. The set of all ports will be denoted by $\mathcal{P}$. The surface at the interface between the simulation domain and port $p$ will be denoted by $\partial \mathcal{Q} \cap \partial\mathcal{P}_p$.

\begin{figure}[t]
	\centering
	\begin{subfigure}[t]{0.8\linewidth}
		\includegraphics[width=\textwidth]{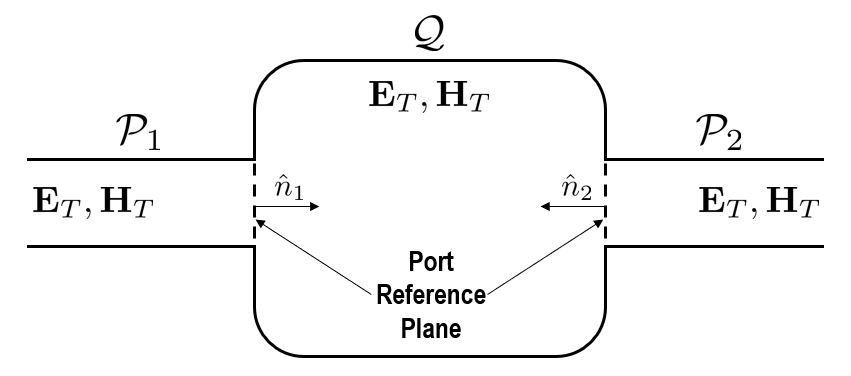}
		\caption{}
		\label{subfig:projector-quantization-setup1}
	\end{subfigure}
	\begin{subfigure}[t]{0.8\linewidth}
		\includegraphics[width=\textwidth]{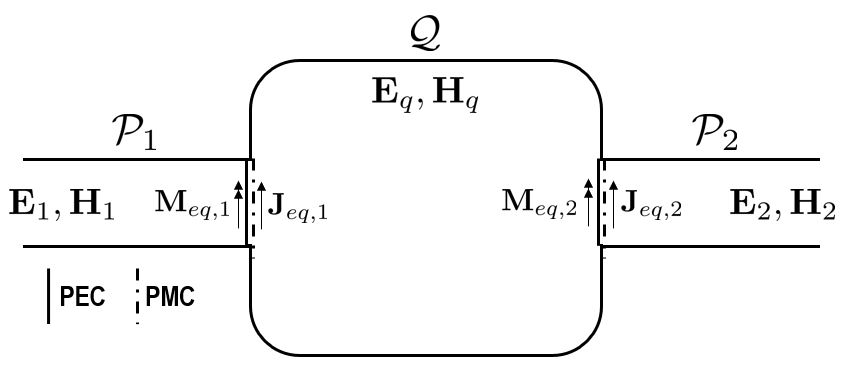}
		\caption{}
		\label{subfig:projector-quantization-setup2}
	\end{subfigure}
	\caption{Illustration of the projector-based quantization setup. In (a) an example two-port problem is shown, while in (b) artificial boundaries and equivalent currents are introduced to separate the regions of the problem.}
	\label{fig:projector-quantization-setup}
\end{figure}

We now present a physically-motivated development of this quantization approach based on an analysis method for open cavities \cite[Ch. 2.9]{haus2012electromagnetic}. This approach also has similarities with using waveports in various computational electromagnetics methods \cite{wang2015higher}. To help guide the discussion, an illustration of the problem setup for this quantization approach is shown in Fig. \ref{fig:projector-quantization-setup}. In Fig. \ref{subfig:projector-quantization-setup1}, the original problem is shown with two reference planes for ports identified. The true fields in all regions of the problem are $\mathbf{E}_T$ and $\mathbf{H}_T$. Now, the regions of the problem are separated by introducing perfect electric conductor (PEC) or perfect magnetic conductor (PMC) boundary conditions in the simulation domain at all port interfaces, as shown in Fig. \ref{subfig:projector-quantization-setup2}. To maintain the hermiticity of the entire problem, complementary conditions are used to close the port region problems \cite{viviescas2003field}. That is, if a PMC condition closes the simulation domain the corresponding port is closed with a PEC condition, as shown in Fig. \ref{subfig:projector-quantization-setup2}.

The artificial ``closing'' surfaces lead to discontinuities in the electric or magnetic fields at these locations that should not be present from the original problem shown in Fig. \ref{subfig:projector-quantization-setup1}. To produce the correct fields within the simulation domain, equivalent electric or magnetic current densities are introduced at the closing surfaces. For instance, if a PMC condition is applied at a region of the simulation domain (c.f. $\partial\mathcal{Q}\cap\partial\mathcal{P}_1$ in Fig. \ref{subfig:projector-quantization-setup2}), the resulting discontinuity in the magnetic field is compensated with an equivalent electric current density given by $\mathbf{J}_{eq,p} = \hat{n}_p\times \mathbf{H}_T$. Here, $\hat{n}_p$ points into the simulation domain and $\mathbf{H}_T$ should be expanded in terms of the port modes to tie the two problems together \cite{haus2012electromagnetic}. Similarly, to produce the correct fields within the corresponding port region, an equivalent magnetic current density given by $\mathbf{M}_{eq,p} = \hat{n}_p\times \mathbf{E}_T$ must be introduced in the port region (c.f. $\partial\mathcal{Q}\cap\partial\mathcal{P}_1$ in Fig. \ref{subfig:projector-quantization-setup2}). Here, $\mathbf{E}_T$ should be expanded in terms of the simulation domain modes and the unusual sign in the definition of $\mathbf{M}_{eq,p}$ is due to the fixed polarity of the unit normal vector $\hat{n}_p$.

From this physical picture, we see that the interaction between the simulation domain and port regions can be achieved by introducing equivalent current densities. Considering, for now, only closing the simulation domain with PMC conditions, a $\mathbf{J}_{eq,p}$ will need to be introduced at each port in the simulation domain. In Lagrangian/Hamiltonian treatments of electromagnetics, the interaction between a current $\mathbf{J}$ and the field is typically given in terms of the vector potential as $\mathbf{A}\cdot\mathbf{J}$ \cite{chew2016quantum2}. Hence, our resulting Hamiltonian should be
\begin{multline}
H_F = \frac{1}{2}\iiint \big( \epsilon |\mathbf{E}_q|^2 + \mu | \mathbf{H}_q|^2 + \sum_{p \in \mathcal{P}} \big[ \epsilon |\mathbf{E}_p|^2  \\ + \mu | \mathbf{H}_p|^2  \big]  - \sum_{p \in \mathcal{P}} 2 \mathbf{A}_q \cdot (\hat{n}_p\times \mathbf{H}_p ) \big) d\mathbf{r},
\label{eq:interacting-system-hamiltonian}
\end{multline} 
where a subscript of $q$ ($p$) denotes that the term is associated with the simulation domain (ports). Further, the term $\hat{n}_p \times \mathbf{H}_p$ is an equivalent electric current density (with $\hat{n}_p$ pointing into the simulation domain). The more general case involving both artificial PEC and PMC conditions will be handled in Section \ref{sec:eom}, where it will also be shown that this Hamiltonian produces the correct equations of motion (i.e., Maxwell's equations fed by electric and magnetic current sources).

Inspecting the coupling term in (\ref{eq:interacting-system-hamiltonian}), we see that it has been written from the perspective of treating the port fields as a source to the simulation domain. It is of course possible to also look at the Hamiltonian from the alternative viewpoint that the ports are being fed by a current density. This is done by rearranging the coupling term to be $\mathbf{H}_p\cdot(\mathbf{A}_q\times\hat{n}_p)$, which shows the magnetic field coupling to a term that is proportional to a $\mathbf{M}_{eq,p}$. Although difficult to see at this point, this coupling term will produce the correct form of Maxwell's equations with a $\mathbf{M}_{eq,p}$ acting as a source to the port field equations. This will be shown in Section \ref{sec:eom}.

With the Hamiltonian formulated, the modal expansion of the fields needs to be revisited. By construction of the problem, a complete set of modes can be found in each region to expand the fields in a piecewise manner. Hence, we have that the simulation domain electric field is
\begin{align}
\mathbf{E}_q(\mathbf{r},t) =  \sum_k \sqrt{\frac{\omega_k}{2\epsilon_0}} \big( q_k(t) \mathbf{E}_k(\mathbf{r}) + q^*_k(t) \mathbf{E}^*_k(\mathbf{r})  \big)
\label{eq:sim-e-field}
\end{align}
and the port region electric fields are
\begin{multline}
\mathbf{E}_p(\mathbf{r},t) = \sum_\lambda \int_0^\infty d\omega_{\lambda p}\, \sqrt{\frac{\omega_{\lambda p}}{2\epsilon_0}} \big( q_{\lambda p}(\omega_{\lambda p},t) \mathbf{E}_{\lambda p}(\mathbf{r},\omega_{\lambda p}) \\ + q^*_{\lambda p}(\omega_{\lambda p},t) \mathbf{E}^*_{\lambda p}(\mathbf{r},\omega_{\lambda p})    \big).
\label{eq:port-e-field}
\end{multline}
In (\ref{eq:sim-e-field}), the summation over $k$ represents a discrete spectrum of modes with eigenvalue $\omega_k$ for the region $\mathcal{Q}$. In (\ref{eq:port-e-field}), the index $p$ is used to differentiate the different ports in the set $\mathcal{P}$. Each port can support different transverse modes (e.g., transverse electromagnetic or transverse electric), which are differentiated by the discrete index $\lambda$. Due to the semi-infinite length of the port regions, each transverse mode will also support a continuous spectrum. Hence, the integration over the eigenvalue $\omega_{\lambda p}$ can be interpreted as ``continuously summing'' over the one-dimensional continuum of modes for each transverse mode. The overall fields are then the summation of the various mode expansions, i.e., $\mathbf{E} = \mathbf{E}_q + \sum_{p\in\mathcal{P}} \mathbf{E}_p$.

A similar expansion also holds for the magnetic field, i.e., $\mathbf{H} = \mathbf{H}_q + \sum_{p\in\mathcal{P}} \mathbf{H}_p$ where
\begin{align}
\mathbf{H}_q(\mathbf{r},t) = \sum_k \sqrt{\frac{\omega_k}{2\mu_0}} \big( p_k(t) \mathbf{H}_k(\mathbf{r}) + p^*_k(t) \mathbf{H}^*_k(\mathbf{r})  \big),
\end{align}
\begin{multline}
\mathbf{H}_p(\mathbf{r},t) = \sum_\lambda \int_0^\infty d\omega_{\lambda p}\, \sqrt{\frac{\omega_{\lambda p}}{2\mu_0}} \big( p_{\lambda p}(\omega_{\lambda p},t) \mathbf{H}_{\lambda p}(\mathbf{r},\omega_{\lambda p}) \\ + p^*_{\lambda p}(\omega_{\lambda p},t) \mathbf{H}^*_{\lambda p}(\mathbf{r},\omega_{\lambda p})    \big).
\end{multline}
As suggested by the Hamiltonian, an expansion for $\mathbf{A}$ is also necessary. To be consistent with the the expansions of $\mathbf{E}$ and $\mathbf{H}$, the modal expansions for $\mathbf{A}$ are
\begin{align}
\mathbf{A}_q(\mathbf{r},t) = -i\sum_k \sqrt{\frac{1}{2\omega_k\epsilon_0}} \big( p_k(t) \mathbf{E}_k(\mathbf{r})  - p^*_k(t) \mathbf{E}^*_k(\mathbf{r})  \big),
\end{align}
\begin{multline}
\mathbf{A}_p(\mathbf{r},t) = -i\sum_\lambda \int_0^\infty d\omega_{\lambda p}\, \sqrt{\frac{1}{2\omega_{\lambda p}\epsilon_0}} \times \\ \big( p_{\lambda p}(\omega_{\lambda p},t) \mathbf{E}_{\lambda p}(\mathbf{r}, \omega_{\lambda p})  - p^*_{\lambda p}(\omega_{\lambda p},t) \mathbf{E}^*_{\lambda p}(\mathbf{r},\omega_{\lambda p})    \big),
\end{multline}
with $\mathbf{A} = \mathbf{A}_q + \sum_{p\in\mathcal{P}} \mathbf{A}_p$. To simplify the derivation, we use the radiation gauge defined by $\nabla\cdot\epsilon\mathbf{A} = 0, \Phi = 0$ in this work. This gauge is valid when there are either no near-field sources or when only transverse currents are considered, i.e., $\nabla\cdot\mathbf{J}=0$ \cite{jackson1999classical}.

These modal expansions can now be substituted into the Hamiltonian given in (\ref{eq:interacting-system-hamiltonian}). This gives
\begin{align}
H_F = H_\mathcal{Q} + H_\mathcal{P} + H_{\mathcal{QP}},
\label{eq:proj-hamiltonian-sho}
\end{align}
where
\begin{align}
H_\mathcal{Q} = \sum_k \frac{\omega_k}{2}\big( |q_k(t)|^2 + |p_k(t)|^2  \big),
\end{align}
\begin{multline}
H_\mathcal{P} =  \sum_{p\in\mathcal{P}} \sum_\lambda \int_0^\infty d\omega_{\lambda p} \, \frac{\omega_{\lambda p} }{2} \big( |q_{\lambda p}(\omega_{\lambda p},t)|^2 \\ + |p_{\lambda p}(\omega_{\lambda p},t)|^2 \big),
\end{multline}
\begin{multline}
H_{\mathcal{QP}} =  \sum_{p\in\mathcal{P}} \sum_{\lambda, k}  \int_0^\infty d\omega_{\lambda p} \,  \big( \mathcal{W}_{k,\lambda p}(\omega_{\lambda p}) p^*_k(t) p_{\lambda p}(\omega_{\lambda p},t) \\ + \mathcal{V}_{k, \lambda p}(\omega_{\lambda p}) p_k(t) p_{\lambda p}(\omega_{\lambda p},t)  + \mathrm{H.c.}  \big),
\label{eq:proj-hamiltonian-sho2}
\end{multline}
\begin{align}
\mathcal{W}_{k,\lambda p}(\omega_{\lambda p}) = - \mathcal{K}_{k,\lambda p} \int \mathbf{E}^*_k(\mathbf{r})  \cdot \hat{n}_p\!\times\!\nabla\!\times\!\mathbf{E}_{ \lambda p}(\mathbf{r}, \omega_{\lambda p}) dS ,
\label{eq:coupling1}
\end{align}
\begin{align}
\mathcal{V}_{k,\lambda p}(\omega_{\lambda p}) = \mathcal{K}_{k, \lambda p} \int \mathbf{E}_k(\mathbf{r}) \cdot \hat{n}_p\!\times\!\nabla\!\times\!\mathbf{E}_{\lambda p}(\mathbf{r},\omega_{\lambda p}) dS,
\label{eq:coupling2}
\end{align}
\begin{align}
\mathcal{K}_{k,\lambda p} = \frac{c_0^2}{2}\sqrt{\frac{1}{\omega_k\omega_{\lambda p}}}.
\end{align}
In (\ref{eq:coupling1}) and (\ref{eq:coupling2}), the integration surface is $\partial \mathcal{Q} \cap \partial\mathcal{P}_p$ and the magnetic field has been rewritten in terms of the vector potential to more closely match the formulas given in \cite{viviescas2003field}. Inspecting (\ref{eq:proj-hamiltonian-sho}) to (\ref{eq:proj-hamiltonian-sho2}), $H_\mathcal{Q}$ and $H_\mathcal{P}$ represent summations of simple harmonic oscillators for each mode of the fields, while $H_{\mathcal{QP}}$ represents coupling between these oscillators. Further, considering that $\hat{n}_p\!\times\!\nabla\!\times\!\mathbf{E}_{\lambda p}$  is proportional to electric field modes \cite[Ch. 2]{haus2012electromagnetic}, the coupling terms given in (\ref{eq:coupling1}) and (\ref{eq:coupling2}) are proportional to overlap integrals of different spatial mode profiles. These overlap integrals will weight how strongly the harmonic oscillators from the different regions interact, which is physically intuitive from a mode matching perspective. 

The form of the Hamiltonian given in (\ref{eq:proj-hamiltonian-sho}) suggests the quantization process \cite{viviescas2003field}. Similar to the modes-of-the-universe case, each mode can be quantized by elevating the harmonic oscillator variables to be quantum operators with equal-time commutation relations 
\begin{align}
[\hat{q}_{k_1}(t),\hat{p}_{k_2}(t)] = i\hbar \delta_{k_1,k_2},
\end{align}
\begin{multline}
[\hat{q}_{\lambda_1 p_1}(\omega_{\lambda_1 p_1},t),\hat{p}_{\lambda_2 p_2}(\omega_{\lambda_2 p_2}',t)] = i\hbar \delta_{\lambda_1, \lambda_2} \delta_{p_1,p_2} \\ \times \delta(\omega_{\lambda_1 p_1}-\omega_{\lambda_2 p_2}').
\end{multline}
In addition to these commutation relations, the operators from different regions commute with each other.

Now, bosonic ladder operators can be introduced for each mode similar to (\ref{eq:annihilation}) to (\ref{eq:boson-commutation}). Using these, the total electric field operator is $\hat{\mathbf{E}} = \hat{\mathbf{E}}_q + \sum_{p\in\mathcal{P}}\hat{\mathbf{E}}_p$ where
\begin{align}
\hat{\mathbf{E}}_q(\mathbf{r},t) =  \sum_k N_{E,k} \big( \hat{a}_k(t) \mathbf{u}_k(\mathbf{r}) + \hat{a}^\dagger_k(t) \mathbf{u}^*_k(\mathbf{r})    \big),
\label{eq:q-sim-e-field}
\end{align}
\begin{multline}
\hat{\mathbf{E}}_p(\mathbf{r},t) = \sum_\lambda \int_0^\infty d\omega_{\lambda p} \, N_{E,\lambda p} \big( \hat{a}_{\lambda p}(\omega_{\lambda p},t) \mathbf{u}_{\lambda p}(\mathbf{r},\omega_{\lambda p})   \\ +  \hat{a}^\dagger_{\lambda p}(\omega_{\lambda p},t) \mathbf{u}^*_{\lambda p}(\mathbf{r},\omega_{\lambda p})     \big).
\label{eq:q-port-e-field}
\end{multline}
A similar expansion holds for the magnetic field operator, i.e., $\hat{\mathbf{H}} = \hat{\mathbf{H}}_q + \sum_{p\in\mathcal{P}}\hat{\mathbf{H}}_p$ where
\begin{align}
\hat{\mathbf{H}}_q(\mathbf{r},t) = \sum_k N_{H,k} \big( \hat{a}_k(t) \mathbf{v}_k(\mathbf{r}) + \hat{a}^\dagger_k(t) \mathbf{v}^*_k(\mathbf{r})    \big),
\label{eq:q-sim-h-field}
\end{align}
\begin{multline}
\hat{\mathbf{H}}_p(\mathbf{r},t) = \sum_\lambda \int_0^\infty d\omega_p \, N_{H,\lambda p} \big( \hat{a}_{\lambda p}(\omega_{\lambda p},t) \mathbf{v}_{\lambda p}(\mathbf{r},\omega_{\lambda p})  \\ +  \hat{a}^\dagger_{\lambda p}(\omega_{\lambda p},t) \mathbf{v}^*_{\lambda p}(\mathbf{r},\omega_{\lambda p})     \big) .
\label{eq:q-port-h-field}
\end{multline}
Further, we have that $\hat{\mathbf{A}} = \hat{\mathbf{A}}_q + \sum_{p\in\mathcal{P}}\hat{\mathbf{A}}_p$ where
\begin{align}
\hat{\mathbf{A}}_q(\mathbf{r},t) =  -i\sum_k N_{A,k} \big( \hat{a}_k(t) \mathbf{u}_k(\mathbf{r}) - \hat{a}^\dagger_k(t) \mathbf{u}^*_k(\mathbf{r})    \big),
\label{eq:q-sim-a-field}
\end{align}
\begin{multline}
\hat{\mathbf{A}}_p(\mathbf{r},t)\! = \!-i\sum_\lambda \int_0^\infty d\omega_{\lambda p } \, N_{A,\lambda p} \big( \hat{a}_{\lambda p}(\omega_{\lambda p},t) \mathbf{u}_{\lambda p}(\mathbf{r},\omega_{\lambda p})   \\ -  \hat{a}^\dagger_{\lambda p}(\omega_{\lambda p},t) \mathbf{u}^*_{\lambda p}(\mathbf{r},\omega_{\lambda p})     \big),
\label{eq:q-port-a-field}
\end{multline}
\begin{align}
N_{A,k(\lambda p)} = \sqrt{\frac{\hbar }{2\epsilon_0 \omega_{k(\lambda p)} N_{E,k(\lambda p)} }}.
\end{align}
In (\ref{eq:q-sim-e-field}) to (\ref{eq:q-port-a-field}), the modal representations of $\mathbf{u}_k$, $\mathbf{u}_p$, $\mathbf{v}_k$, and $\mathbf{v}_p$ have been used to allow the modal normalizations to be explicitly included in the field operators. This is useful when establishing a correspondence between Hamiltonians written in terms of field and transmission line operators.

The corresponding Hamiltonian is
\begin{multline}
\hat{H}_F = \frac{1}{2}\iiint \big( \epsilon \hat{\mathbf{E}}^2_q + \mu \hat{\mathbf{H}}^2_q + \sum_{p \in \mathcal{P}} \big[ \epsilon \hat{\mathbf{E}}^2_p + \mu \hat{\mathbf{H}}^2_p  \big] \\ - \sum_{p \in \mathcal{P}} 2 \hat{\mathbf{A}}_q \cdot ( \hat{n}_p \times \hat{\mathbf{H}}_p ) \big) d\mathbf{r}.
\label{eq:q-interacting-system-hamiltonian}
\end{multline}
This can be expressed using bosonic ladder operators as 
\begin{align}
\hat{H}_F = \hat{H}_\mathcal{Q} + \hat{H}_\mathcal{P} + \hat{H}_{\mathcal{Q}\mathcal{P}},
\label{eq:q-interacting-system-hamiltonian2}
\end{align}
where
\begin{align}
\hat{H}_\mathcal{Q} = \sum_k \hbar \omega_k \hat{a}^\dagger_k (t) \hat{a}_k(t)
\end{align}
\begin{align}
\hat{H}_\mathcal{P} = \sum_{p\in \mathcal{P}} \sum_\lambda \int_0^\infty d\omega_{\lambda p} \, \hbar \omega_{\lambda p} \hat{a}^\dagger_{\lambda p}(\omega_{\lambda p},t) \hat{a}_{\lambda p}(\omega_{\lambda p},t) 
\end{align}
\begin{multline}
\hat{H}_{\mathcal{Q}\mathcal{P}} =  \sum_{p\in \mathcal{P}} \sum_{\lambda, k} \int_0^\infty d\omega_{\lambda p} \big( \mathcal{W}_{k,\lambda p}(\omega_{\lambda p}) \hat{a}^\dagger_k(t) \hat{a}_{\lambda p}(\omega_{\lambda p},t) \\ + \mathcal{V}_{k,\lambda p}(\omega_{\lambda p}) \hat{a}_k(t) \hat{a}_{\lambda p}(\omega_{\lambda p},t) + \mathrm{H.c.}   \big).
\end{multline}
The Hamiltonian given in (\ref{eq:q-interacting-system-hamiltonian2}) can be recognized as the system-and-bath Hamiltonian that is commonly used in quantum optics \cite{viviescas2003field,gardiner2004quantum}. This is a satisfying result, since this Hamiltonian is often used to study the input-output relationship of optical cavities \cite{walls2007quantum}.

%% file: field-tr-correspondence2.tex
\section{Correspondence Between Field and Transmission Line Hamiltonians}
\label{sec:field-to-circuit}
With an appropriate quantization procedure now in place, we can continue the process of developing a field-based description of circuit QED systems. To assist in this, it will be useful to determine a correspondence between the field-based Hamiltonian of (\ref{eq:free-field-qH}) or (\ref{eq:q-interacting-system-hamiltonian}) and a Hamiltonian consisting of transmission line voltages and currents. To simplify this process, the classical case from the modes-of-the-universe approach; i.e., (\ref{eq:free-field-cH}), is considered first. 

Since our goal is to reduce our expressions to a form like the circuit-based description given in (\ref{eq:coupled-transmon}), we need to introduce some assumptions implicit in (\ref{eq:coupled-transmon}) for the manipulations in this section and Section \ref{sec:field-transmon-hamiltonian} to work. However, we emphasize that these approximations are only needed to show consistency with the approximate expressions used in the literature. They are not needed in the construction of our general field-based equations provided up to this point.
	
Now, the first assumption is that our transmission line geometry and operating frequencies are such that only quasi-TEM modes are excited. For simplicity, these modes will be treated as pure TEM modes for the purposes of defining transmission line parameters such as voltages, currents, and per-unit-length impedances. Second, we will assume in this section that we are only considering a finite length transmission line with a constant cross-section and that fringing effects at the end of the transmission line can be accounted for separately (e.g., by shifting mode frequencies). For notational simplicity, the longitudinal direction of the transmission line will be aligned with the $z$-axis. Considering these simplifications, we can arrive at an ``exact'' correspondence between field-based and transmission line-based descriptions in this section.

To begin, we need to revisit the expansion of the electric and magnetic fields in terms of modes. Since we are dealing with TEM waves, we can decompose the electric field as 
\begin{multline}
\mathbf{E}(\mathbf{r},t) = \sum_{k,l}  \sqrt{\frac{\omega_{k,l}}{2\epsilon_0 N_{E_{T,k}} N_{E_{L,l}} }} \\ \times \big( q_{k,l}(t) \mathbf{u}_{k,l}(\mathbf{r})  +  q^*_{k,l}(t) \mathbf{u}^*_{k,l}(\mathbf{r})  \big)
\label{eq:emode}
\end{multline}
where the mode functions $\mathbf{u}_{k,l}$ are split into a transverse vector function and a longitudinal scalar function as
\begin{align}
\mathbf{u}_{k,l}(\mathbf{r}) = \mathbf{u}_{T,k}(x,y) u_{L,l}(z).
\end{align}
These mode functions are orthogonal in the sense that
\begin{align}
\iint \epsilon_r\mathbf{u}_{T,k_1}^* \cdot \mathbf{u}_{T,k_2} dxdy = \delta_{k_1,k_2} N_{E_{T,k_1}},
\label{eq:norm1}
\end{align}
\begin{align}
\int u_{L,l_1}^*(z) u_{L,l_2}(z) dz = \delta_{l_1,l_2} N_{E_{L,l_1}}.
\label{eq:norm2}
\end{align}
Similarly, for the magnetic field we have that
\begin{multline}
\mathbf{H}(\mathbf{r},t) = \sum_{k,l} \sqrt{\frac{\omega_{k,l}}{2\mu_0 N_{H_{T,k}} N_{H_{L,l}} }} \\ \times   \big( p_{k,l}(t) \mathbf{v}_{k,l}(\mathbf{r})  +  p^*_{k,l}(t) \mathbf{v}^*_{k,l}(\mathbf{r})  \big)
\end{multline}
where the mode functions $\mathbf{v}_{k,l}$ are
\begin{align}
\mathbf{v}_{k,l}(\mathbf{r}) = \mathbf{v}_{T,k}(x,y) v_{L,l}(z).
\end{align}
These mode functions are orthogonal in a similar sense to (\ref{eq:norm1}) and (\ref{eq:norm2}).

The conversion between fields and transmission line quantities can be performed by adopting definitions for the transmission line parameters so that the power and energy densities expressed in terms of field and circuit parameters agree \cite{pozar2009microwave}. Since we consider lossless lines here, only the per-unit-length capacitance and inductance of the line are needed. For a particular transmission line mode, these are denoted as $C_k$ and $L_k$, respectively. Within the notation of this work, the definitions for these become $C_k = \epsilon_0 N_{E_{T,k}}$ and $L_k = \mu_0 N_{H_{T,k}}$ \cite{pozar2009microwave}.

With these definitions, the field-Hamiltonian of (\ref{eq:free-field-cH}) can now be converted into a transmission line form. The electric field term will be considered first. Substituting in the modal expansion, this term becomes
\begin{multline}
\epsilon |\mathbf{E}(\mathbf{r},t)|^2 = \epsilon \! \sum_{k_1,l_1,k_2,l_2}\!\!\sqrt{\frac{\omega_{k_1,l_1}\omega_{k_2,l_2}}{\epsilon^2_0 N_{E_{T,k_1}} N_{E_{L,l_1}} N_{E_{T,k_2}} N_{E_{L,l_2}} } } \\ \times \big(   q^*_{k_2,l_2}(t) q_{k_1,l_1}(t) \mathbf{u}^*_{k_2,l_2}(\mathbf{r}) \cdot \mathbf{u}_{k_1,l_1}(\mathbf{r}) \\ + \frac{1}{2} q_{k_2,l_2}(t)q_{k_1,l_1}(t) \mathbf{u}_{k_2,l_2}(\mathbf{r})  \cdot \mathbf{u}_{k_1,l_1}(\mathbf{r}) \\ + \frac{1}{2} q^*_{k_2,l_2}(t)q^*_{k_1,l_1}(t) \mathbf{u}^*_{k_2,l_2}(\mathbf{r})  \cdot \mathbf{u}^*_{k_1,l_1}(\mathbf{r}) \big) .
\label{eq:e1}
\end{multline}
Recalling that the Hamiltonian involves the volume integral of these terms, we can simplify our expression by noting that the terms proportional to $\mathbf{u}_{k_2,l_2}\cdot\mathbf{u}_{k_1,l_1}$ and $(\mathbf{u}_{k_2,l_2}\cdot\mathbf{u}_{k_1,l_1})^*$ will average to zero unless $k_1=k_2$ and $l_1=l_2$. We then also have that for harmonic oscillators, terms of the form $q_{k,l}^2$ and $(q_{k,l}^*)^2$ vanish \cite{haken1976quantum}. Hence, we have that
\begin{multline}
\!\!\iiint \epsilon |\mathbf{E}(\mathbf{r},t)|^2 d\mathbf{r} = \\ \!\!\! \sum_{k_1,l_1,k_2,l_2} \sqrt{\frac{\omega_{k_1,l_1}\omega_{k_2,l_2}}{\epsilon^2_0 N_{E_{T,k_1}} N_{E_{L,l_1}} N_{E_{T,k_2}} N_{E_{L,l_2}} } }  \\ \times q^*_{k_2,l_2}(t) q_{k_1,l_1}(t) \iiint \epsilon \,   \mathbf{u}^*_{k_2,l_2}(\mathbf{r}) \cdot \mathbf{u}_{k_1,l_1}(\mathbf{r}).
\end{multline}
We can expand the $\mathbf{u}_{k,l}$ functions and use the orthogonality relationships given in (\ref{eq:norm1}) and (\ref{eq:norm2}) to get
\begin{multline}
\iiint \epsilon |\mathbf{E}(\mathbf{r},t)|^2d\mathbf{r} \\ = \int \sum_{k,l} C_k  \frac{\omega_{k,l}}{C_k N_{E_{L,l}}} |q_{k,l}(t)|^2 |u_{L,l}(z)|^2 dz,
\label{eq:int1}
\end{multline}
where we have also noted that $C_k = \epsilon_0 N_{E_{T,k}}$. We don't cancel the $C_k$ terms in (\ref{eq:int1}) because it helps suggest the correct form of voltage mode to convert between the electric field and transmission line voltage descriptions of the system.

In particular, we can define a voltage mode to be
\begin{align}
V_{k,l}(z,t) = \sqrt{\frac{\omega_{k,l}}{ C_k N_{E_{L,l}}}} \big(q_{k,l}(t) u_{L,l}(z) + q_{k,l}^*(t) u_{L,l}^*(z)  \big).
\end{align}
Following a similar set of steps to those shown in (\ref{eq:e1}) to (\ref{eq:int1}), we can see that
\begin{align}
\iiint \epsilon |\mathbf{E}(\mathbf{r},t)|^2d\mathbf{r} = \int \sum_{k,l} C_k |V_{k,l}(z,t)|^2 dz.
\label{eq:eint}
\end{align}
Similarly, We can define a current mode as
\begin{align}
I_{k,l}(z,t) = \sqrt{\frac{\omega_{k,l}}{ L_k N_{H_{L,l}}}} \big(p_{k,l}(t) v_{L,l}(z) + p_{k,l}^*(t) v_{L,l}^*(z)  \big)
\end{align}
to see that 
\begin{align}
\iiint \mu |\mathbf{H}(\mathbf{r},t)|^2d\mathbf{r} = \int \sum_{k,l} L_k |I_{k,l}(z,t)|^2 dz.
\label{eq:hint}
\end{align}
Combining the results in (\ref{eq:eint}) and (\ref{eq:hint}), the transmission line Hamiltonian can be written as
\begin{align}
H_{TR} = \int \frac{1}{2} \sum_{k,l}  \big( C_k |V_{k,l}(z,t)|^2 + L_k|I_{k,l}(z,t)|^2     \big)dz,
\end{align}
which is equivalent to the field-based Hamiltonian. This equality is of course only valid when considering a portion of a system with a constant transmission line cross-section, as mentioned at the beginning of this section. However, this is exactly the part of a system used in writing a circuit-based Hamiltonian like (\ref{eq:coupled-transmon}). Hence, we now have the tools to relate a field-based Hamiltonian to the simpler circuit descriptions often used in the literature. 

Moving now to the quantum case, it is important to note that all the operations used in the classical case still apply. Hence, the process easily generalizes to the quantum case, giving
\begin{align}
\hat{H}_{F} = \hat{H}_{TR} = \int \frac{1}{2} \sum_{k,l} \big( C_k \hat{V}_{k,l}^2 + L_k \hat{I}_{k,l}^2    \big) dz,
\end{align}
where 
\begin{align}
\hat{V}_{k,l}(z,t) = N_{V_{k,l}} \big( \hat{a}_{k,l}(t) u_{L,l}(z) + \hat{a}_{k,l}^\dagger(t)u^*_{L,l}(z)  \big),  
\label{eq:c-q-voltage}
\end{align}
\begin{align}
\hat{I}_{k,l}(z,t) = N_{I_{k,l}} \big( \hat{a}_{k,l}(t)v_{L,l}(z) + \hat{a}_{k,l}^\dagger(t)v^*_{L,l}(z)  \big) ,
\label{eq:c-q-current}
\end{align}
\begin{align}
N_{V_{k,l}} = \sqrt{ \frac{\hbar \omega_{k,l}}{2 C_k N_{E_{L,l}}} }, \,\, N_{I_{k,l}} = \sqrt{ \frac{\hbar \omega_{k,l}}{2 L_k N_{H_{L,l}}} }.
\label{eq:sim-domain-mode-norm}
\end{align}

These definitions for $\hat{V}_{k,l}$ and $\hat{I}_{k,l}$ are not immediately seen to be consistent with those given in (\ref{eq:total-v}) and (\ref{eq:total-i}). The reason is that complex-valued spatial modes have been used in (\ref{eq:c-q-voltage}) and (\ref{eq:c-q-current}), while real-valued modes were used in (\ref{eq:total-v}) and (\ref{eq:total-i}). Further, the spatial variation of the voltage and current modes have been completely integrated out for (\ref{eq:total-v}) and (\ref{eq:total-i}).

To see that the expressions derived in this section are consistent with the literature, the steps outlined in this section can be repeated for the real-valued spatial mode expansions like those given in (\ref{eq:r-q-efield}) and (\ref{eq:r-q-hfield}). For this case, the voltage and current operators become
\begin{align}
\hat{V}_{k,l}(z,t) =  N_{V_{k,l}} \big( \hat{a}_{k,l}(t)  + \hat{a}_{k,l}^\dagger(t) \big)  u_{L,l}(z)
\label{eq:r-q-voltage}
\end{align}
\begin{align}
\hat{I}_{k,l}(z,t) = -i N_{I_{k,l}} \big( \hat{a}_{k,l}(t) - \hat{a}_{k,l}^\dagger(t)  \big) v_{L,l}(z).
\label{eq:r-q-current}
\end{align}
These can be seen to be consistent with (\ref{eq:total-v}) and (\ref{eq:total-i}) by recalling that $C_r = C\ell$ and $L_r = L\ell$, where $\ell$ is the length of the resonator \cite{blais2004cavity}. Restricting (\ref{eq:r-q-voltage}) and (\ref{eq:r-q-current}) to be a resonator mode, it can be seen that the longitudinal normalizations $N_{E_{L,k}}$ and $N_{H_{L,k}}$ would become $\ell$. The additional factors that occur after integrating out the remaining spatial variation in (\ref{eq:r-q-voltage}) and (\ref{eq:r-q-current}) are grouped with other terms in the overall Hamiltonian to be consistent with \cite{blais2004cavity}.

With the basic process developed, the expressions from the projector-based quantization approach, e.g., (\ref{eq:q-interacting-system-hamiltonian}), can now be converted into a transmission line form as well. This will not be performed here for brevity. However, it should be noted that the interacting part of the Hamiltonian given in (\ref{eq:q-interacting-system-hamiltonian}) cannot be written in a simpler transmission line form. This is because the transverse integrations cannot be concisely written when allowing for complex-valued mode functions. If real-valued mode functions are used, the spatial integrals can be shown to be proportional to overlap integrals of the electric field modes for the different regions of the problem, which is an intuitively satisfying result.

%% file: field-based-transmon2.tex
\section{Hamiltonian for the Field-Transmon System}
\label{sec:field-transmon-hamiltonian}
With the ability to convert between field and transmission line representations, it is now possible to show the consistency of the postulated field-based Hamiltonian of (\ref{eq:field-transmon-hamiltonian1}) with the more common circuit-based description given in (\ref{eq:coupled-transmon}). To do this, we must match the assumptions implicit in (\ref{eq:coupled-transmon}), which correspond to only considering the unperturbed TEM modes of a transmission line resonator that interact with the transmon qubit. This approximately corresponds to only considering the portion of Fig. \ref{fig:region-illustration} marked with length $\ell$. To simplify the expressions, only fields from the simulation domain will be considered as these are what interact with the transmon. The subscript of $q$ will be dropped from these fields in this section. Further, all of the simulation domain mode functions will be assumed to be real-valued. Since the simulation domain can typically be made a closed system, this choice does not amount to a loss of generality \cite{chew2016quantum2}. The complex-valued mode function case can be handled easily, but leads to unnecessarily long expressions that are omitted for brevity.

Considering these points, the postulated field-transmon system Hamiltonian of (\ref{eq:field-transmon-hamiltonian1}) is reproduced here in full for ease of reference as
\begin{multline}
\hat{H} =  4E_C (\hat{n}-n_g)^2 - E_J \cos \hat{\varphi} \\ + \iiint \frac{1}{2} \big( \epsilon \hat{\mathbf{E}}^2 + \mu \hat{\mathbf{H}}^2 - 2  \hat{\mathbf{E}} \cdot \partial_t^{-1} \hat{\mathbf{J}}_t  \big)d\mathbf{r},
\label{eq:field-transmon-hamiltonian}
\end{multline}
where the transmon current density operator was
\begin{align}
\hat{\mathbf{J}}_t = -2e \beta \mathbf{d} \delta(z-z_0) \partial_t\hat{n} .
\label{eq:transmon-current-operator}
\end{align}
The field-based Hamiltonian of (\ref{eq:field-transmon-hamiltonian}) can be shown to be equivalent to circuit-based Hamiltonians by evaluating the spatial integration in (\ref{eq:field-transmon-hamiltonian}). The first two terms can be simply evaluated following the steps in Section \ref{sec:field-to-circuit}, yielding
\begin{multline}
\iiint \frac{1}{2} \big( \epsilon \hat{\mathbf{E}}^2 + \mu \hat{\mathbf{H}}^2 \big) d\mathbf{r} \\ = \frac{1}{2}\sum_{k,l} \big(  C_k N_{E_{L,l}} \hat{V}^2_{I_{k,l}} + L_k N_{H_{L,l}} \hat{I}^2_{I_{k,l}}   \big),
\label{eq:f-to-tr1}
\end{multline}
where the integrated voltage and current operators are 
\begin{align}
\hat{V}_{I_{k,l}} = N_{V_{k,l}} \big( \hat{a}_{k,l}(t) + \hat{a}_{k,l}^\dagger(t)  \big)
\label{eq:q-total-voltage-mode}
\end{align}
\begin{align}
\hat{I}_{I_{k,l}} = -i N_{I_{k,l}} \big( \hat{a}_{k,l}(t) - \hat{a}_{k,l}^\dagger(t)  \big).
\label{eq:q-total-current-mode}
\end{align}

The remaining term to consider is the coupling term. To carry out the spatial integration, the expression is expanded in terms of the modes for $\hat{\mathbf{E}}$ given in (\ref{eq:emode}). This gives
\begin{multline}
-\iiint \hat{\mathbf{E}}(\mathbf{r},t) \cdot \partial_t^{-1} \hat{\mathbf{J}}_t(\mathbf{r},t) d\mathbf{r} = 
\sum_{k,l} N_{E_{k,l}} \big( \hat{a}_{k,l}(t) + \hat{a}^\dagger_{k,l}(t) \big)  \\ \times 2e \beta \hat{n}(t) \iiint \mathbf{u}_{k,l}(\mathbf{r}) \cdot \mathbf{d}(x,y) \delta(z-z_0) d\mathbf{r},
\end{multline}
where
\begin{align}
N_{E_{k,l}} = \sqrt{ \frac{\hbar \omega_{k,l}}{2 \epsilon_0 N_{E_{T,k}} N_{E_{L,l}}} }.
\label{eq:enorm}
\end{align}
Now, the spatial integral along the $z$-axis can be evaluated easily and the remaining transverse integral can be identified as the definition of a voltage, i.e.,
\begin{align}
V_{k,l}(z_0) = \int_{a}^{b} \mathbf{u}_{k,l}(x,y,z_0) \cdot d \mathbf{d}(x,y),
\end{align}
where $a$ and $b$ are the initial and final points of the integration path defined by $\mathbf{d}$ \cite{pozar2009microwave}. Hence, we have that
\begin{multline}
-\iiint \hat{\mathbf{E}}(\mathbf{r},t) \cdot \partial_t^{-1} \hat{\mathbf{J}}_t(\mathbf{r},t) d\mathbf{r} \\ = 
2e \beta \hat{n}(t) \sum_{k,l} N_{V_{k,l}} \big( \hat{a}_{k,l}(t) + \hat{a}^\dagger_{k,l}(t) \big)  V_{k,l}(z_0)  ,
\label{eq:int3}
\end{multline}
where we have rewritten $N_{E_{k,l}}$ as $N_{V_{k,l}}$ by noting the relationship between the transverse normalization in (\ref{eq:enorm}) and the per-unit-length modal capacitance in (\ref{eq:sim-domain-mode-norm}). Finally, we can use (\ref{eq:q-total-voltage-mode}) to get
\begin{multline}
-\iiint \hat{\mathbf{E}}(\mathbf{r},t) \!\cdot\! \partial_t^{-1} \hat{\mathbf{J}}_t(\mathbf{r},t) d\mathbf{r} \\  = \sum_{k,l} 2e \beta V_{k,l}(z_0) \hat{V}_{I_{k,l}}(t) \hat{n}(t).
\label{eq:coupling-derivation}
\end{multline}

Putting the results of (\ref{eq:f-to-tr1}) and (\ref{eq:coupling-derivation}) together, the field-transmon system Hamiltonian of (\ref{eq:field-transmon-hamiltonian}), can now be written as
\begin{multline}
\hat{H} =  4E_C (\hat{n}-n_g)^2 - E_J \cos \hat{\varphi} +  \frac{1}{2}\sum_{k,l} \big(  C_k N_{E_{L,l}} \hat{V}^2_{I_{k,l}} \\ + L_k N_{H_{L,l}} \hat{I}^2_{I_{k,l}}   \big) + \sum_{k,l} 2e \beta V_{k,l}(z_0) \hat{V}_{I_{k,l}} \hat{n}.
\end{multline}
Restricting this Hamiltonian to only consider a single mode of a resonator coupled to the transmon recovers (\ref{eq:coupled-transmon}). Hence, the postulated field-transmon system Hamiltonian can be seen to be consistent with the circuit-based descriptions of circuit QED systems typically used in the literature.

%% file: eom.tex
\section{Equations of Motion}
\label{sec:eom}
Now that an appropriate Hamiltonian has been found for the field-transmon system, the quantum equations of motion can be derived using Hamilton's equations \cite{chew2016quantum}. We will consider the full system composed of the transmon qubit, the simulation domain fields, and the port region fields. The Hamiltonian is also generalized by allowing for the termination of the port regions in the simulation domain with either PEC or PMC conditions. 

With this understood, the complete Hamiltonian becomes
\begin{align}
\hat{H} = \hat{H}_T + \hat{H}_F + \hat{H}_I ,
\label{eq:quantum-hamiltonian}
\end{align}
where
\begin{align}
\hat{H}_T = 4E_C (\hat{n}-n_g)^2 - E_J \cos \hat{\varphi},
\label{eq:transmon-hamiltonian}
\end{align}
\begin{align}
\hat{H}_F = \frac{1}{2}\iiint \big( \epsilon \hat{\mathbf{E}}^2_q + \mu \hat{\mathbf{H}}^2_q  + \sum_{p \in \mathcal{P}} \big[ \epsilon \hat{\mathbf{E}}^2_p + \mu \hat{\mathbf{H}}^2_p  \big]  \big) d\mathbf{r} ,
\label{eq:field-hamiltonian}
\end{align}
\begin{multline}
\hat{H}_I = - \iiint \big( \hat{\mathbf{E}}_q \cdot \partial_t^{-1}\hat{\mathbf{J}}_t  + \sum_{p \in \mathcal{P}_M}  \hat{\mathbf{A}}_q \cdot (\hat{n}_p\times \hat{\mathbf{H}}_p)  \\ + \sum_{p \in \mathcal{P}_E}  \hat{\mathbf{F}}_q \cdot (\hat{\mathbf{E}}_p \times \hat{n}_p) \big) d\mathbf{r}.
\label{eq:interaction-hamiltonian}
\end{multline}
In (\ref{eq:interaction-hamiltonian}), $\mathcal{P}_E$ ($\mathcal{P}_M$) denotes the set of ports terminated in a PEC (PMC) \textit{in the simulation domain}. The union of these sets is all of the ports $\mathcal{P}$. Note that $\hat{n}_p$ is the unit normal vector to the port surface, and it points into the simulation domain. The terms in (\ref{eq:interaction-hamiltonian}) quantify the interactions between the different parts of the total system. The first term is the coupling of the simulation domain field and transmon system, while the next two terms represent coupling between the simulation domain and port region fields.

In (\ref{eq:interaction-hamiltonian}), the electric vector potential, $\hat{\mathbf{F}}_q$, has been introduced into the Hamiltonian. This is necessary because of the set of ports $\mathcal{P}_E$ that introduce equivalent magnetic current densities as sources to the simulation domain. The simplest way to account for the presence of magnetic sources is to introduce another set of auxiliary potentials, as is commonly done in classical electromagnetics \cite{jin2011theory}.

To derive equations of motion, the Hamiltonian needs to be expressed in terms of canonical conjugate operators \cite{chew2016quantum}. The transmon operators $\hat{n}$ and $\hat{\varphi}$ are already in this form. However, the electric and magnetic fields are not canonical conjugate operators in a Hamiltonian mechanics formalism \cite{chew2016quantum}. Instead, the electromagnetic field portions of the Hamiltonian need to be rewritten in terms of the electric and magnetic vector potentials and their conjugate momenta.

To support this, the electric and magnetic fields are first decomposed into the set of fields produced by electric or magnetic sources. Under this decomposition, the fields produced by electric (magnetic) sources are completely specified by the magnetic (electric) vector potential \cite{jin2011theory}. For this to hold, a radiation gauge is being used for both the magnetic and electric vector potentials. Considering this, the field portions of the Hamiltonian are first rewritten as
\begin{multline}
\hat{H}_F = \frac{1}{2} \iiint \big( \epsilon\hat{\mathbf{E}}_{qe}^2 + \mu\hat{\mathbf{H}}_{qe}^2 + \epsilon\hat{\mathbf{E}}_{qm}^2 + \mu\hat{\mathbf{H}}_{qm}^2 \\ + \sum_{p \in \mathcal{P}} \big[ \epsilon \hat{\mathbf{E}}^2_{pe} + \mu \hat{\mathbf{H}}^2_{pe} + \epsilon \hat{\mathbf{E}}^2_{pm} + \mu \hat{\mathbf{H}}^2_{pm}\big]   \big) d\mathbf{r}  ,
\end{multline}
\begin{multline}
\hat{H}_I = -\iiint \big(  \hat{\mathbf{E}}_{qe} \cdot  \partial_t^{-1} \hat{\mathbf{J}}_t  + \sum_{p \in \mathcal{P}_M}  \hat{\mathbf{A}}_{qe} \cdot (\hat{n}_p\times \hat{\mathbf{H}}_{pm})  \\ + \sum_{p \in \mathcal{P}_E}  \hat{\mathbf{F}}_{qm} \cdot (\hat{\mathbf{E}}_{pe} \times \hat{n}_{p}) \big) d\mathbf{r},
\end{multline}
where a subscript $e$ ($m$) denotes that this quantity is due to electric (magnetic) sources. The structure of the coupling terms between the fields in different regions reflects the difference in boundary conditions and corresponding equivalent source densities at the interfaces between regions.

With the Hamiltonian decomposed into portions due to electric and magnetic sources, it can now be written in terms of the electric and magnetic vector potentials and their conjugate momenta. This gives
\begin{multline}
\hat{H}_F = \frac{1}{2} \iiint \big( \epsilon^{-1}\hat{\mathbf{\Pi}}_{qe}^2 + \mu^{-1}(\nabla\times\hat{\mathbf{A}}_{qe})^2 + \mu^{-1}\hat{\mathbf{\Pi}}_{qm}^2 \\ + \epsilon^{-1}(\nabla\times\hat{\mathbf{F}}_{qm})^2   + \sum_{p \in \mathcal{P}} \big[ \epsilon^{-1}\hat{\mathbf{\Pi}}_{pe}^2 + \mu^{-1}(\nabla\times\hat{\mathbf{A}}_{pe})^2  \\ + \mu^{-1}\hat{\mathbf{\Pi}}_{pm}^2 + \epsilon^{-1}(\nabla\times\hat{\mathbf{F}}_{pm})^2\big]  \big) d\mathbf{r},
\end{multline}
\begin{multline}
\hat{H}_I =  \iiint \big( \epsilon^{-1}\hat{\mathbf{\Pi}}_{qe} \cdot \partial_t^{-1} \hat{\mathbf{J}}_t  - \! \sum_{p \in \mathcal{P}_M}  \!\!\mu^{-1} \hat{\mathbf{A}}_{qe} \!\cdot\! (\hat{n}_p \! \times \! \hat{\mathbf{\Pi}}_{pm})  \\ - \sum_{p \in \mathcal{P}_E}  \epsilon^{-1} \hat{\mathbf{F}}_{qm} \cdot (\hat{\mathbf{\Pi}}_{pe} \times \hat{n}_{p}) \big) d\mathbf{r},
\end{multline}
where $\hat{\mathbf{\Pi}}_{qe} = \epsilon\partial_t \hat{\mathbf{A}}_{qe}$ is the conjugate momentum for the vector potential in the simulation domain. Similarly, $\hat{\mathbf{\Pi}}_{pe} = \epsilon\partial_t \hat{\mathbf{A}}_{pe}$ is the conjugate momentum for the vector potential in the port regions. The conjugate momenta for the electric vector potentials are $\hat{\mathbf{\Pi}}_{qm} = \mu \partial_t \hat{\mathbf{F}}_{qm}$ and $\hat{\mathbf{\Pi}}_{pm} = \mu \partial_t \hat{\mathbf{F}}_{pm}$ for the simulation domain and port regions, respectively.

With the Hamiltonian now written completely in terms of canonical conjugate operators, equations of motion can be derived using Hamilton's equations \cite{chew2016quantum}. Equations of motion will first be derived for the transmon operators. For these operators, Hamilton's equations are 
\begin{align}
\frac{\partial \hat{\varphi}}{\partial t} = \frac{\partial \hat{H}}{\partial \hat{n}} , \,\, \frac{\partial \hat{n}}{\partial t} = -\frac{\partial \hat{H}}{\partial \hat{\varphi}} .
\end{align}
Evaluating the necessary derivatives gives
\begin{multline}
\frac{\partial \hat{\varphi}}{\partial t} = 8 E_C (\hat{n}-n_g) \\ + 2e\beta \iiint \hat{\mathbf{E}}_{qe} (\mathbf{r},t) \cdot \mathbf{d} \delta(z-z_0) d\mathbf{r},
\end{multline}
\begin{align}
\frac{\partial \hat{n}}{\partial t} = - E_J \sin\hat{\varphi},
\end{align}
where the field-transmon interaction term has been rewritten in terms of the electric field operator. 

Equations of motion for the potentials requires taking functional derivatives of $\hat{H}$ with respect to the conjugate operators \cite{chew2016quantum}. These can be easily performed, and will be done in stages for the different sets of potentials. Beginning with the simulation domain magnetic vector potential, we have
\begin{align}
\frac{\delta\hat{H}}{\delta \hat{\mathbf{A}}_{qe}} = \mu^{-1} \nabla\times\nabla\times\hat{\mathbf{A}}_{qe}   +  \sum_{p \in \mathcal{P}_M}\mu^{-1} \hat{n}_p\times \hat{\mathbf{\Pi}}_{pm},
\end{align}
\begin{align}
\frac{\delta\hat{H}}{\delta \hat{\mathbf{\Pi}}_{qe}} =  \epsilon^{-1} \hat{\mathbf{\Pi}}_{qe}  + \epsilon^{-1} \partial_t^{-1} \hat{\mathbf{J}}_t .
\end{align}
Hamilton's equations of motion for the magnetic vector potential system are \cite{chew2016quantum}
\begin{align}
\frac{\partial \hat{\mathbf{A}}_{qe}}{\partial t} = \frac{\delta\hat{H}}{\delta \hat{\mathbf{\Pi}}_{qe}}, \,\, \frac{\partial \hat{\mathbf{\Pi}}_{qe}}{\partial t} = -\frac{\delta\hat{H}}{\delta \hat{\mathbf{A}}_{qe}}.
\label{eq:ham1}
\end{align}
These can be combined to give
\begin{align}
\nabla\times\nabla\times\hat{\mathbf{A}}_{qe} + \mu\epsilon \partial_t^2 \hat{\mathbf{A}}_{qe} =  \mu \hat{\mathbf{J}}_t - \sum_{p \in \mathcal{P}_M} \hat{n}_p \times \hat{\mathbf{\Pi}}_{pm}.
\label{eq:sim-A-wave}
\end{align}
This inhomogeneous wave equation can be seen to take the expected form for the magnetic vector potential by noting that $-\hat{n}_p\times\hat{\mathbf{\Pi}}_{pm} = \mu \hat{n}_p\times\hat{\mathbf{H}}_{pm}$ is an equivalent electric current density times the permeability. 

A similar process can be done for the magnetic vector potential in the port regions. For a particular port, we have
\begin{align}
\frac{\delta\hat{H}}{\delta \hat{\mathbf{A}}_{pe}} = \mu^{-1} \nabla\times\nabla\times\hat{\mathbf{A}}_{pe} 
\end{align}
\begin{align}
\frac{\delta\hat{H}}{\delta \hat{\mathbf{\Pi}}_{pe}} =  \epsilon^{-1} \hat{\mathbf{\Pi}}_{pe} + \sum_{p' \in \mathcal{P}_E} \delta_{p,p'}\epsilon^{-1} \hat{n}_{p'} \times\hat{\mathbf{F}}_{qm}.
\end{align}
The Kronecker delta function is used to only include a source term if the particular port $p \in \mathcal{P}_E$. Similar functional derivatives can be evaluated for each of the individual port regions. Hamilton's equations can then be used to derive a wave equation. This yields
\begin{align}
\nabla\times\nabla\times\hat{\mathbf{A}}_{pe} + \mu\epsilon \partial^2_t \hat{\mathbf{A}}_{pe} = \!\!\sum_{p' \in \mathcal{P}_E} \delta_{p,p'} \mu \hat{n}_{p'} \! \times\!\partial_t\hat{\mathbf{F}}_{qm}.
\label{eq:port-A-wave}
\end{align}
Note that due to the fixed orientation of $\hat{n}_p$ pointing into the simulation domain, $\hat{n}_p \times\partial_t \hat{\mathbf{F}}_{qm}$ is an equivalent electric current density with a positive sign.

To finish the derivation, equations of motion for the electric vector potential need to be established. As expected, these follow a very similar process to that for the magnetic vector potential. Beginning with the equations for the simulation domain, the necessary functional derivatives are
\begin{align}
\frac{\delta\hat{H}}{\delta \hat{\mathbf{F}}_{qm}} =  \nabla\times\epsilon^{-1}\nabla\times\hat{\mathbf{F}}_{qm} -  \sum_{p \in \mathcal{P}_E}\epsilon^{-1} \hat{n}_p\times \hat{\mathbf{\Pi}}_{pm}
\end{align}
\begin{align}
\frac{\delta\hat{H}}{\delta \hat{\mathbf{\Pi}}_{qm}} =  \mu^{-1} \hat{\mathbf{\Pi}}_{qm}.
\end{align}
Hamilton's equations of motion for the electric vector potential system are
\begin{align}
\frac{\partial \hat{\mathbf{F}}_{qm}}{\partial t} = \frac{\delta\hat{H}}{\delta \hat{\mathbf{\Pi}}_{qm}}, \,\, \frac{\partial \hat{\mathbf{\Pi}}_{qm}}{\partial t} = -\frac{\delta\hat{H}}{\delta \hat{\mathbf{F}}_{qm}}.
\label{eq:ham3}
\end{align}
These can be combined to give
\begin{align}
\epsilon\nabla\times\epsilon^{-1}\nabla\times\hat{\mathbf{F}}_{qm} + \mu\epsilon \partial_t^2 \hat{\mathbf{F}}_{qm} =   \sum_{p \in \mathcal{P}_M} \hat{n}_p \times \hat{\mathbf{\Pi}}_{pm}.
\label{eq:sim-F-wave}
\end{align}
By recalling that $\hat{n}_p \times\hat{\mathbf{\Pi}}_{pm} = \epsilon \hat{\mathbf{E}}_{pm} \times\hat{n}_p$, it is seen that the source term for this inhomogeneous wave equation has the form of an equivalent magnetic current density times the permittivity. Hence, this is the expected wave equation for an electric vector potential in the radiation gauge.

The final set of equations are for the electric vector potential in the port regions. The functional derivatives are
\begin{align}
\frac{\delta\hat{H}}{\delta \hat{\mathbf{F}}_{pm}} =  \nabla\times\epsilon^{-1}\nabla\times\hat{\mathbf{F}}_{pm} 
\end{align}
\begin{align}
\frac{\delta\hat{H}}{\delta \hat{\mathbf{\Pi}}_{pm}} =  \mu^{-1} \hat{\mathbf{\Pi}}_{pm} - \sum_{p' \in \mathcal{P}_M} \delta_{p,p'}\mu^{-1} \hat{n}_{p'} \times\hat{\mathbf{A}}_{qe}.
\end{align}
Using Hamilton's equations, the results of these functional derivatives can be combined to give
\begin{multline}
\epsilon\nabla\times\epsilon^{-1}\nabla\times\hat{\mathbf{F}}_{pm} + \mu\epsilon \partial^2_t \hat{\mathbf{F}}_{pm} \\ = -\sum_{p' \in \mathcal{P}_M} \delta_{p,p'} \epsilon \hat{n}_{p'} \times\partial_t\hat{\mathbf{A}}_{qe}.
\label{eq:port-F-wave}
\end{multline}
Similar to (\ref{eq:port-A-wave}), the fixed polarity of $\hat{n}_p$ means that $-\hat{n}_p\times\partial_t\hat{\mathbf{A}}_{qe}$ is equal to an equivalent magnetic current density with a positive sign.

Noting that (\ref{eq:sim-A-wave}), (\ref{eq:port-A-wave}), (\ref{eq:sim-F-wave}), and (\ref{eq:port-F-wave}) are the expected wave equations in each region for the radiation gauge, it can be concluded that the equations of motion for the electromagnetic fields are simply the quantum Maxwell's equations for each region with the necessary sources added \cite{chew2016quantum2}. For the simulation domain, this gives
\begin{multline}
\nabla\times\hat{\mathbf{H}}_q(\mathbf{r},t) - \partial_t \hat{\mathbf{D}}_q(\mathbf{r},t) \\ = \hat{\mathbf{J}}_t(\mathbf{r},t) + \sum_{p \in \mathcal{P}_M} \hat{\mathbf{J}}_{p}(\mathbf{r},t) 
\end{multline}
\begin{align}
\nabla\times\hat{\mathbf{E}}_q(\mathbf{r},t) + \partial_t \hat{\mathbf{B}}_q(\mathbf{r},t) = - \sum_{p \in \mathcal{P}_E} \hat{\mathbf{M}}_{p}(\mathbf{r},t) 
\end{align}
\begin{align}
\nabla \cdot \hat{\mathbf{D}}_q(\mathbf{r},t) = 0
\end{align}
\begin{align}
 \nabla \cdot \hat{\mathbf{B}}_q(\mathbf{r},t) = 0
\end{align}
where the port current densities are
\begin{align}
\hat{\mathbf{J}}_p(\mathbf{r},t) = \hat{n}_p \times \hat{\mathbf{H}}_p(\mathbf{r},t), \,\, \,\,\,\,\, p \in \mathcal{P}_M, 
\end{align}
\begin{align}
\hat{\mathbf{M}}_p(\mathbf{r},t) = -\hat{n}_p \times \hat{\mathbf{E}}_p(\mathbf{r},t), \,\, \,\,\,\,\, p \in \mathcal{P}_E.
\end{align}
Similarly, the quantum Maxwell's equations for a single port region are
\begin{align}
\nabla\times\hat{\mathbf{H}}_p(\mathbf{r},t) - \partial_t \hat{\mathbf{D}}_p(\mathbf{r},t) = \hat{\mathbf{J}}_{q}(\mathbf{r},t)
\end{align}
\begin{align}
\nabla\times\hat{\mathbf{E}}_p(\mathbf{r},t) + \partial_t \hat{\mathbf{B}}_p(\mathbf{r},t) = -  \hat{\mathbf{M}}_{q}(\mathbf{r},t)
\end{align}
\begin{align}
\nabla \cdot \hat{\mathbf{D}}_p(\mathbf{r},t) =  0
\end{align}
\begin{align}
\nabla \cdot \hat{\mathbf{B}}_p(\mathbf{r},t) =  0
\end{align}
where the port current densities are
\begin{align}
\hat{\mathbf{J}}_q(\mathbf{r},t) = -\hat{n}_p \times \hat{\mathbf{H}}_q(\mathbf{r},t), \,\, \,\,\,\,\, p \in \mathcal{P}_E, 
\end{align}
\begin{align}
\hat{\mathbf{M}}_q(\mathbf{r},t) = \hat{n}_p \times \hat{\mathbf{E}}_q(\mathbf{r},t), \,\, \,\,\,\,\, p \in \mathcal{P}_M.
\end{align}
Note that due to the radiation gauge used in this work, the port current densities are all solenoidal and as a result there are no quantum charge densities associated with these currents.

Maxwell's equations can be used to form wave equations for $\hat{\mathbf{E}}$. In the simulation domain, this gives
\begin{multline}
\nabla\times\nabla\times\hat{\mathbf{E}}_q(\mathbf{r},t) + \mu\epsilon\partial_t^2\hat{\mathbf{E}}_q(\mathbf{r},t)  = -\mu \partial_t \hat{\mathbf{J}}_t(\mathbf{r},t) \\ -\mu  \sum_{p\in\mathcal{P}_M} \partial_t \hat{\mathbf{J}}_p(\mathbf{r},t) -  \sum_{p\in\mathcal{P}_E} \nabla\times \hat{\mathbf{M}}_p(\mathbf{r},t),
\label{eq:q-Sim-wave}
\end{multline}
while in a particular port region we have
\begin{multline}
\nabla\times\nabla\times\hat{\mathbf{E}}_p(\mathbf{r},t) + \mu\epsilon\partial_t^2\hat{\mathbf{E}}_p(\mathbf{r},t) \\ =  -\mu  \partial_t \hat{\mathbf{J}}_q(\mathbf{r},t)   -  \nabla\times \hat{\mathbf{M}}_q(\mathbf{r},t).
\label{eq:q-Port-wave}
\end{multline}
In general, only one set of sources will be present in (\ref{eq:q-Port-wave}) depending on whether $p\in \mathcal{P}_E$ or $p\in\mathcal{P}_M$.

With appropriate wave equations developed, different modeling strategies can be devised to solve them. For instance, a quantum finite-difference time-domain solver could be used \cite{na2020quantum2}. Alternatively, eigenmodes of the electromagnetic system can be found numerically and used in a quantum information preserving numerical framework \cite{na2020quantum}. In certain cases, the current densities can be treated as impressed sources so that a dyadic Green's function approach can be used to propagate the quantum information \cite{chew2016quantum2}. We will demonstrate this process for a circuit QED system in our future work. 

Although full-wave numerical models are of primary interest for practical applications, the development of simpler problems that can be solved using analytical methods are also important. The solutions to these test problems can help validate different full-wave modeling techniques and serve as a useful pedagogical tool for learning how this new formalism can be applied to real-world problems.

%% file: conclusion.tex
\section{Conclusion}
\label{sec:conclusion}
In this work, we have provided a new look at how circuit QED systems using transmon qubits can be described mathematically. Expressed in terms of three-dimensional vector fields, this new approach is well-suited to developing numerical models that can leverage the latest developments in computational electromagnetics research. We have also demonstrated how our new model is consistent with the simpler circuit-based descriptions often used in the literature. Using our new model, we derived the quantum equations of motion applicable to the coupled field-transmon system. Developing solution strategies for this kind of coupled quantum system is an area of active research interest. Numerical methods in this area have the potential to greatly benefit the overall field of circuit QED, and correspondingly, the development of new kinds of quantum information processing hardware. 